\documentclass[cmp,final]{svjour}  
\usepackage{amsmath}
\usepackage{amsfonts,amssymb}

\journalname{Communications in Mathematical Physics}

\newcommand{\RM}{\mathbb{R}}

\newcommand{\CM}{\mathbb{C}}

\title{On the Magnetization of a Charged Bose Gas \\ 
in the Canonical Ensemble}
\titlerunning{On the Magnetization of a Charged Bose Gas in 
the Canonical Ensemble}
\author{Horia D. Cornean} 
\institute{Institute of Mathematics of the Romanian Academy, 
P.O. Box 1-764, \\ 70700 Bucharest, Romania; \\ 
\email{hcornean@imar.ro  ;  cornean@barutu.fizica.unibuc.ro}} 
\authorrunning{H. D. Cornean}
\date{Received: 15 July 1999 / Accepted: 29 November 1999}
\communicated{H. Araki}

\begin{document}
\maketitle
\begin{abstract}
Consider a charged 
Bose gas without self-interactions, 
confined in a three dimensional cubic box of side 
$L\geq 1$ and subjected to a constant magnetic field $B\neq 0$. 
If the bulk density of particles $\rho$ and the 
temperature $T$ are fixed, then define the canonical magnetization 
as the partial derivative 
with respect to $B$ of the reduced free energy. Our main 
result is that it admits 
thermodynamic limit for all strictly positive $\rho$, $T$ and $B$. 
It is also proven that the canonical and grand canonical 
magnetizations (the last one at fixed average density) 
are equal up to the surface 
order corrections.
\end{abstract}
\section{Introduction}
\setcounter{equation}{0}
Much work has been done on the thermodynamic behavior of large 
systems composed from independent quantum particles in the presence of 
external magnetic fields. As is 
well known, the fundamental problem consists in 
proving the existence of the thermodynamic limit for the potentials and 
the equations of state defined at finite volume. In the particular 
case of the canonical magnetization (defined 
as the partial derivative with respect to 
the magnetic field of the reduced free energy), one has to prove that 
the derivative (performed at finite volume) 
commutes with the thermodynamic limit of the reduced free 
energy. Although the quantum canonical 
ensemble is (from the physical point of view) the most important one, 
most of the previous works were carried out either using the 
Maxwell-Boltzmann statistics or in the framework 
of the quantum grand canonical ensemble, because in those settings, 
many physically relevant quantities can be expressed employing 
the integral kernel of 
the Gibbs semigroup associated to the one particle problem. Moreover, one is 
able to go beyond the bulk terms and investigate finite size 
effects. 

Take for example 
the grand canonical pressure of a quantum gas in a constant magnetic field. 
The rigorous proof of its thermodynamic limit goes back at least to 
Angelescu and Corciovei [A, A-C]; its surface correction (in the regime 
in which the fugacity is less than one) was obtained by Kunz [K]. As for the 
Maxwell-Boltzmann magnetization, 
nice results were obtained by Macris {\it et al} [M-M-P 1,2]; they 
wrote down even the corner corrections. Notice that in these papers the 
domain $\Lambda$ was allowed to be more general, typically convex with 
piecewise smooth boundary. 

Another result concerning 
the thermodynamic limit and the surface corrections for the magnetization 
and susceptibility of a Fermi gas at zero magnetic field was obtained 
by Angelescu {\it et al} [A-B-N 2]. Because this paper motivated our 
work, we are giving some more details about it. 

Firstly, as in our setting, their domain was 
a rectangular parallelepiped and the 
magnetic field oriented after the third direction. They defined the 
grand canonical magnetization $m_\Lambda(\beta,z)$ 
(susceptibility $\chi_\Lambda(\beta,z)$) as the first (second) 
derivative with respect to the magnetic field of the grand canonical 
pressure at $B=0$, 
for all $z\in {\CM}\setminus (-\infty,-1]$. Their 
main result can be roughly stated as follows: 

i. $m_\Lambda(\beta,z)=0$, $\forall z\in {\CM}\setminus (-\infty,-1]$;

ii. There exists $\chi_\infty(\beta,z)$ analytic in 
${\CM}\setminus (-\infty,-1]$
such that for any compact $K\subset {\CM}\setminus (-\infty,-1]$ one has:

$$\lim_{\Lambda\rightarrow\infty}
\sup_{z\in K}|\chi_\Lambda(\beta,z)- \chi_\infty(\beta,z)|
=0.$$

More than that, they gave even the surface correction for susceptibility 
and proved that this expansion is 
uniform on compacts. Because the relation between the 
fugacity and the grand canonical average density of Fermi 
particles can be always inverted, they 
were able to express the grand canonical susceptibility in terms of the 
canonical parameters $\rho$ and $\beta$. Let us stress that $B=0$ and 
$\Lambda$ a rectangular 
parallelepiped were crucial ingredients in [A-B-N 2], 
the uniform convergence on compacts being obtained via a substantial use of 
the explicit formula of the integral kernel of the Gibbs 
semigroup associated to the Dirichlet Laplacian. 

In this paper, we are studying the ``true'' canonical problem for a Bose gas 
at nonzero magnetic field $B_0 >0$ (in order to avoid the Bose condensation). 
Using a standard procedure (see [K-U-Z, H]) 
of deriving the canonical partition function 
from the grand canonical pressure (see (\ref{cortez2})), we are 
able to transform the uniform convergence on compacts of the grand canonical 
magnetization (see Lemma \ref{limgama}) into a pointwise 
convergence ($\beta$, $\rho$ fixed and $L\rightarrow\infty$) 
of the canonical magnetization; this result is given in 
Theorem \ref{magcano}. Moreover, we obtain that the canonical magnetization 
$m_L$ (see (\ref{cortez4})) and the grand canonical magnetization 
at fixed average density (see (\ref{cortez5})) are equal up to the surface 
order corrections. 

Two natural questions arise: what about Fermi statistics and what about 
higher derivatives with respect to $B$ (the susceptibility for example)? 
Partial answers and a few open problems are outlined at the end of the proofs.

\section{Preliminaries and the Results}
\setcounter{equation}{0}

Let $\Lambda =\left \{ {\bf x}\in {\RM}^3 |\: 
-\frac{L}{2}<x_j<\frac{L}{2},\:j\in\{1,2,3\} \right \}$, $L>1$, be a cubic 
box with 
its side equal to $L$. Then the ``one particle'' Hilbert space is 
${\cal H}_{1,L}:={\bf L}^2(\Lambda)$; denote with 
${\cal H}_{n,L}$ the proper subspace of 
$\otimes_{j=1}^n {\cal H}_{1,L} \cong {\bf L}^2(\Lambda^n)$ 
which contains all totally symmetric functions. Denote with 
${\cal H}_{0,L}={\CM}$ the space with no particles; then the Fock space is 
defined as  
${\cal F}_L :=\bigoplus_{n\geq 0}{\cal H}_{n,L}$. 
One can introduce 
the ``number of particles'' operator $N_L$ as the unique self-adjoint 
extension of the multiplication with $n$ on each ${\cal H}_{n,L}$. 

Assume that the particles (each having an electric charge $e$) 
are subjected to a constant magnetic field 
${\bf B}=B{\bf e_3}$, which corresponds to a magnetic vector potential 
$B{\bf a}=\frac{B}{2}{\bf e_3}\wedge {\bf x}$. If $c$ stands for the speed of 
light, define $\omega :=(e/c)B$. Then the ``one particle'' 
Hamiltonian (denoted with $H_{1,L}(\omega)$) will be the Friederichs extension 
of the symmetric and positive operator 
$\frac{1}{2}\left (-\mbox{ i}\nabla -\omega\: {\bf a}\right )^2$ defined on 
$C_0^\infty (\Lambda)$. 

Due to the regularity of $\Lambda$, $H_{1,L}(\omega)$ is essentially 
self-adjoint on 
$$D=\left \{ f\in C^2 (\Lambda)\cap C^1 (\overline{\Lambda}),\:
f |_{\partial \Lambda}=0,\;\Delta f\in {\bf L}^2(\Lambda)  \right \}.$$ 

The Hamiltonian which describes $n$ particles reads as:
\begin{equation}\label{1.10}
H_{n,L}(\omega)=\underbrace
{H_{1,L}(\omega)\otimes \dots \otimes I +\dots +I\otimes\dots 
\otimes  H_{1,L}(\omega)}_{\mbox{``n'' terms}}.
\end{equation}

The second quantized Hamiltonian $H_{L}(\omega)$ 
is defined as the unique self-adjoint operator on ${\cal F}_L$ 
whose restrictions to ${\cal H}_{n,L}$ coincide with $H_{n,L}(\omega)$. 

If $T>0$ stands for the temperature and $\mu \in {\RM}$ for the chemical 
potential, then define $\beta =\frac{1}{k_B\:T}>0$ and $z=\exp{(\beta\:\mu)}$ 
(the fugacity), where $k_B$ is the Boltzmann  constant. When working in the 
canonical ensemble, one considers that the bulk density of particles $\rho$ 
is constant, therefore the number of particles is defined as 
$N(L):=\rho L^3$. 

As is well known (see [R-S 4]), $H_{1,L}(\omega)$ is positive, unbounded 
and has compact 
resolvent; these imply that its spectrum is purely discrete with 
accumulation point 
at infinity. Moreover, from the min-max principle it follows: 
\begin{equation}\label{minmax}
\inf\sigma (H_{1,L}(\omega))\geq 
\inf\sigma (H_{1,\infty}(\omega))
=\frac{\omega}{2}.
\end{equation}
It is also known that the semigroup 
$W_L(\beta,\omega):=\exp{(-\beta H_{1,L}(\omega))}$ 
is trace class and admits an integral 
kernel $G_{\omega,L}({\bf x},{\bf x'};\beta)$, which is continuous in both its 
``spatial'' variables. The diamagnetic inequality at finite volume (see 
[B-H-L]) reads as:
\begin{equation}\label{1.11}
|G_{\omega,L}({\bf x},{\bf x'};\beta)|\leq G_{0,L}({\bf x},{\bf x'};\beta)\leq 
\frac{1}{(2\pi\beta)^{3/2}}\exp{\left ( -\frac{|{\bf x}-{\bf x'}|^2}
{2\beta}\right)}.
\end{equation}   
If ${\cal I}_1({\bf L}^2(\Lambda))$ denotes the Banach space of 
trace class operators, it follows that:
\begin{equation}\label{1.12}
||W_L(\beta,\omega)||_{{\cal I}_1}= 
{\rm tr}\:W_L(\beta,\omega)\leq \frac{L^3}{(2\pi\beta)^{3/2}}.
\end{equation}

Denote with $\{E_{j}(\omega)\}_{j\in {\bf N}}$ the set of the eigenvalues of 
$H_{1,L}(\omega)$. 
If $\mu <0$, the grand canonical partition function reads as: 
\begin{equation}\label{1.13}
\Xi_L(\beta,z,\omega)= {\rm tr}_{{\cal F}_L}\exp{[-\beta (H_{L}(\omega)-
\mu N_L)]}= \prod_{j=0}^\infty [1- z\exp{(-\beta 
E_j(\omega))}]^{-1}.
\end{equation}
The canonical partition function of our system is:
\begin{equation}\label{partcan}
Z_L(\beta,\rho,\omega)={\rm tr}_{{\cal H}_{N(L),L}}
\exp{(-\beta H_{N(L),L}(\omega))}.
\end{equation}
The link between them is contained in the following equality:
\begin{equation}\label{1.1}
\Xi_L(\beta,z,\omega)=
\sum_{n=0}^\infty z^n \mbox{tr}_{{\cal H}_{n,L}}\exp{[-\beta H_{n,\Lambda}]}.
\end{equation}
Throughout the entire paper, by $\log z$ we shall understand the logarithm 
function restricted to ${\mathbf C}\setminus (-\infty, 0]$. 

Let ${\cal C}$ be a contour which surrounds the origin, does not intersect the 
cut $[1,\infty)$ but contains the spectrum of the trace class operator $zW_L$, 
where $z\in {\CM}\setminus [\exp{(\beta\omega /2)}, \infty )$. Let $q(\xi)=
\frac{1}{\xi}\log (1-\xi)$ be 
an analytic function in the interior of ${\cal C}$. 
Define the following bounded operator:
\begin{equation}\label{1.14}
 q(zW_L)=\frac{1}{2\pi \imath}\int_{{\cal C}}d\xi\:q(\xi)(\xi -zW_L)^{-1}.
\end{equation}

It is easy to see that $\log (1-zW_L)=zW_L\cdot q(zW_L)$ 
and using (\ref{1.13}) 
one obtains:
\begin{equation}\label{1.15}
\log \Xi_L(\beta,z,\omega)=-{\rm tr}\left ( zW_L\cdot q(zW_L)\right ).
\end{equation}

Employing the above expression, one can easily prove that the grand canonical 
potential (seen as a function of $z$) is analytic in 
${\CM}\setminus [ \exp{(\beta\omega /2)}, \infty  ) $. When 
$|z|<1$, (\ref{1.15}) becomes: 
\begin{equation}\label{1.166}
\log \Xi_L(\beta,z,\omega)  = \sum_{n=1}^{\infty}
\frac{z^n}{n}{\rm tr}\left ( W_L^n\right )
= \sum_{n=1}^{\infty}
\frac{z^n}{n}\left ( \int_{\Lambda}d{\bf x}\:
G_{\omega,L}({\bf x},{\bf x};n\beta)\right ).
\end{equation}

The grand canonical pressure and density are defined as:
\begin{equation}\label{1.2}
P_{L}(\beta,z,\omega):=\frac{1}{\beta L^3}\log \Xi_L(\beta,z,\omega)
=-\frac{1}{\beta L^3}\sum_j \log\left (1-ze^{-\beta E_j(\omega)}\right ),
\end{equation}

and

\begin{equation}\label{1.5}
 \rho_L(\beta,z,\omega) 
:=\beta z \frac{\partial P_L}{\partial z}(\beta,z,\omega).
\end{equation}
Let us remark that $\rho_L(\beta,x,\omega)$ is an increasing function if 
$0<x<\exp{(\beta\omega /2)}$:
\begin{equation}\label{encrease}
\frac{\partial\rho_{L}}{\partial x}(\beta,x,\omega)=
1/L^3 {\rm tr}[(1-xW_L)^{-2}W_L]>0.
\end{equation}
The proof of the thermodynamic limit for these two quantities goes back at 
least to Angelescu and 
Corciovei [A, A-C]. Because this result plays an important role in our 
work, we shall reproduce it here. In order to do that, let us define 
($\omega >0$):
\begin{equation}\label{corc1}
P_{\infty}(\beta,z,\omega):= \omega \frac{1}{(2\pi\beta)^{3/2}}
\sum_{k=0}^\infty g_{3/2}\left (z e^{-(k+1/2)\omega\beta}\right ),
\end{equation}

and 

\begin{equation}\label{corc2}
\rho_{\infty}(\beta,z,\omega):=\beta z
\frac{\partial P_\infty}{\partial z}(\beta,z,\omega)=
\beta\omega\frac{1}{(2\pi\beta)^{3/2}}
\sum_{k=0}^\infty g_{1/2}\left (z e^{-(k+1/2)\omega\beta}\right ),
\end{equation}
where $g_\sigma(\zeta)$ are the usual Bose functions:
\begin{equation}\label{gbosonic}
g_\sigma(\zeta)=\frac{\zeta}{\Gamma(\sigma)}
\int_0^\infty dt\;\frac{t^{\sigma -1}e^{-t}}
{1-\zeta e^{-t}},
\end{equation}
analytic in 
${\CM}\setminus [1, \infty )$ 
and if $|\zeta|<1$, they are given 
by the following expansion:
$$g_\sigma(\zeta)=\sum_{n=1}^\infty\frac{\zeta^n}{n^\sigma}\;.$$

Then the following result is true (see [A, A-C]):
\begin{theorem}\label{ancorc}
Let 
$K\subset {\CM}\setminus [\exp{(\beta\omega /2)}, \infty )$ 
be a compact set. 
Then 
the grand canonical pressure and density admit the thermodynamic limit i.e.:
\begin{equation}\label{corc3}
\lim_{L\rightarrow\infty}\sup_{z\in K}|P_{L}(\beta,z,\omega)-
P_{\infty}(\beta,z,\omega)|=0,
\end{equation}

and

\begin{equation}\label{corc4}
\lim_{L\rightarrow\infty}\sup_{z\in K}|\rho_{L}(\beta,z,\omega)-
\rho_{\infty}(\beta,z,\omega)|=0.
\end{equation}
\end{theorem}

Firstly, because $P_L$ and $\rho_L$ are analytic functions, then via the 
Cauchy  integral formula it follows that all their complex derivatives admit 
a limit which is uniform on compacts. In particular:
\begin{equation}\label{corcio5}
\lim_{L\rightarrow\infty}\sup_{z\in K}\left \arrowvert \frac{\partial\rho_{L}}
{\partial z}(\beta,z,\omega)-\frac{\partial\rho_{\infty}}
{\partial z}(\beta,z,\omega)\right \arrowvert =0.
\end{equation}

It can be seen from (\ref{corc2}) that $\lim_{x\nearrow e^{\beta\omega/2}}
\rho_\infty(\beta,x,\omega)=\infty$, which means that the Bose condensation 
is absent when a nonzero magnetic field is present. A very important 
consequence of the theorem is that the relation between the fugacity and 
density can be inverted for all temperatures and moreover, 
if $0<x_\infty(\beta,\rho,\omega)< e^{\beta\omega/2}$ is the 
unique real and positive solution of the equation 
$\rho_{\infty}(\beta,x,\omega)=\rho$ and if $x_L(\beta,\rho,\omega)$ is the 
unique real and positive solution which solves 
$\rho_{L}(\beta,x,\omega)=\rho$, then $\lim_{L\rightarrow\infty}x_L=x_\infty$. 

Let us perform the Legendre transform at finite volume:
\begin{equation}\label{corcio6}
\tilde{f}_L(\beta,\rho,\omega):=-P_L(\beta,x_L(\beta,\rho,\omega),\omega)+
\frac{\rho}{\beta}\log x_L(\beta,\rho,\omega).
\end{equation}
A straightforward result is that $\tilde{f}_L$ has the following limit:
\begin{equation}\label{corcio7}
f_\infty(\beta,\rho,\omega):=
-P_\infty(\beta,x_\infty(\beta,\rho,\omega),\omega)+
\frac{\rho}{\beta}\log x_\infty(\beta,\rho,\omega).
\end{equation}

Let us denote with $\frac{\partial W_L}{\partial\omega}(\beta,\omega_0)$ 
the following integral which makes sense in the norm topology of 
$B({\bf L}^2(\Lambda))$ (see [A-B-N 1]):
\begin{equation}\label{dewlaom}
-\int_0^\beta d\tau\;W_L(\beta-\tau,\omega_0)
\;[{\bf a}\cdot({\bf p}-\omega_0 {\bf a})]W_L(\tau,\omega_0).
\end{equation}

A particular case of the problem treated in [A-B-N 1] is that 
$\frac{\partial W_L}{\partial\omega}(\beta,\omega_0)$ is even trace class and 
moreover, for $\delta\omega$ sufficiently small one has:
 
\begin{equation}\label{usd}
\left \Arrowvert 
W_L(\beta,\omega_0+\delta\omega)-W_L(\beta,\omega_0)-\delta\omega
\frac{\partial W_L}{\partial\omega}\right \Arrowvert_{{\cal I}_1}
= {\cal O}((\delta\omega)^2).
\end{equation}
Using for $P_L(\beta,z,\omega_0)$ the following expression:
$$-\frac{1}{\beta L^3}\;{\rm tr}[\log(1-zW_L(\beta,\omega_0))],$$
then the estimate (\ref{usd}) justifies the definition of 
the grand canonical magnetization:
\begin{equation}\label{usd'}
\Gamma_L(\beta,z,\omega_0):=
-\frac{e}{c}\;\frac{\partial P_L}{\partial\omega}(\beta,z,\omega_0)=
-\frac{e\;z}{c\beta L^3}\;{\rm tr}
\left [(1-zW_L(\beta,\omega_0))^{-1}
\frac{\partial W_L}{\partial\omega}\right ].
\end{equation}

From its definition, one can easily see that $\Gamma_L$ has the same domain of 
analyticity in $z$. Now let us define the natural candidate for its 
thermodynamic limit:
\begin{equation}\label{gamainfty}
\Gamma_\infty(\beta,z,\omega_0):=
-\frac{e}{c}\; \frac{\partial P_\infty}{\partial\omega}(\beta,z,\omega_0).
\end{equation}

Our main technical result is presented in the following lemma:
\begin{lemma}\label{limgama}
Let 
$K\subset {\CM}\setminus [\exp{(\beta\omega /2)}, \infty )$ 
be a compact set. 
Then 
the grand canonical magnetization admits the thermodynamic limit i.e.:
\begin{equation}\label{cortez}
\lim_{L\rightarrow\infty}\sup_{z\in K}|\Gamma_{L}(\beta,z,\omega)-
\Gamma_{\infty}(\beta,z,\omega)|=0.
\end{equation}
\end{lemma}

Let us go back to the canonical ensemble. From (\ref{1.1}), (\ref{1.2}) 
and (\ref{partcan}), one can 
write down an useful representation of the canonical partition function:
\begin{equation}\label{cortez2}
Z_L(\beta,\rho,\omega)=\frac{1}{2\pi\imath}\int_{C_1}d\xi\;
\frac{1}{\xi}\left [\frac{\exp{\left (\frac{\beta}{\rho} P_L(\beta,\xi,\omega)
\right )}}{\xi}\right ]^{N(L)},
\end{equation}
where $C_1$ is a contour which surrounds the origin and avoids the cut. 
The reduced free energy reads as:
\begin{equation}\label{cortez3}
f_L(\beta,\rho,\omega):=-\frac{1}{\beta L^3}\log Z_L(\beta,\rho,\omega).
\end{equation}
The canonical magnetization is defined as follows:
\begin{equation}\label{cortez4}
m_L(\beta,\rho,\omega):=\frac{e}{c}\frac{\partial f_L}{\partial\omega}
(\beta,\rho,\omega).
\end{equation}
We expect that $m_L$ should be close to the following quantity:
\begin{equation}\label{cortez5}
\tilde{m}_L(\beta,\rho,\omega):=\frac{e}{c}
\frac{\partial \tilde{f}_L}{\partial\omega}
(\beta,\rho,\omega)=\Gamma_L(\beta,x_L(\beta,\rho,\omega),\omega),
\end{equation}
which converges to 
$\Gamma_\infty(\beta,x_\infty(\beta,\rho,\omega),\omega)$. 

We are able now to give our main result:
\begin{theorem}\label{magcano}
Fix $0<\delta <1/2$. 
For all strictly positive temperatures, bulk densities and magnetic fields, 
the canonical magnetization $m_L(\beta,\rho,\omega)$ admits the thermodynamic 
limit. Moreover, there exist two positive 
constants $C_\delta(\beta,\rho,\omega)$ and $L_\delta(\beta,\rho,\omega)$ 
such that for all $L\geq L_\delta$ one has:
\begin{equation}\label{cortez6}
|m_L -\tilde{m}_L|\leq C_\delta \;L^{-3/2+\delta}.
\end{equation}
\end{theorem}
{\it Remark.} It is clear that $\tilde{m}_L$ is a much more convenient 
quantity. If (at least for dilute gases) one would be able to write down an 
expansion for $\tilde{m}_L$ which takes into account the surface 
corrections:
\begin{equation}\label{cortez7}
\tilde{m}_L=m_\infty +\frac{1}{L}m_S +o(1/L),
\end{equation}
then the estimate (\ref{cortez6}) would imply that the same expansion is 
true for $m_L$, too.

\section{The Proof of Theorem \ref{magcano}}
\setcounter{equation}{0}

At this point, we shall consider that Lemma \ref{limgama} is true and give 
its proof in the next section. In order to simplify the notations, 
we shall drop the dependence on $\beta$, $\rho$ 
and $\omega$ but we shall reintroduce it when needed. 

The main idea of the proof consists in isolating the principal part of the 
integral from (\ref{cortez2}). Although this procedure is far from being new 
(in the physical literature it is known as the Darwin-Fowler method; 
see [K-U-Z, H] and references therein), 
we decided to give a rather detailed proof in order to have a clearer image 
of the remainder from (\ref{cortez6}). 
Firstly, let us choose the contour $C_1$ as 
follows:
\begin{equation}\label{cortez8}
C_1:=\{x_L e^{\imath\phi},\;\phi\in [-\pi,\pi]\}.
\end{equation}
Using (\ref{corcio6}), the formula (\ref{cortez2}) can be rewritten as:
\begin{eqnarray}\label{cortez9}
Z_L &=& \exp{(-\beta \tilde{f}_L L^3)}\frac{1}{2\pi}\int_{-\pi}^{\pi} d\phi\;
\exp{\left [\frac{N(L)\beta (P_L(x_L e^{\imath\phi})-P_L(x_L))}{\rho}\right ]}
e^{-\imath N(L)\phi}= \nonumber \\
&=&  \exp{(-\beta \tilde{f}_L L^3)}N(L)^{-1/2}{\cal A}_L,
\end{eqnarray}
where ${\cal A}_L(\beta,\rho,\omega)$ is given by:
\begin{equation}\label{cortez10}
{\cal A}_L=\frac{\sqrt{N(L)}}{2\pi}\int_{-\pi}^{\pi} d\phi\;
e^{\left [N(L)\beta/\rho 
(\Re P_L(x_L e^{\imath\phi})-P_L(x_L))\right ]}
e^{\imath N(L)\left [\beta\Im P_L(x_L e^{\imath\phi})/\rho - \phi\right ]}.
\end{equation}

Denote with $\tilde{p}_\infty(\phi)$ the function given by 
$\Re P_\infty(x_\infty
e^{\imath\phi})$, and with $\tilde{p_L}(\phi)$ the function given by 
$\Re P_L(x_L e^{\imath\phi})$. 

We shall prove that $\frac{d^2\tilde{p}_\infty}{d\phi^2}(0)<0$ and 
that the following limit is true:
\begin{equation}\label{cortez11}
\lim_{L\rightarrow\infty}{\cal A}_L:={\cal A}_\infty
=\sqrt{-\frac{\rho}{2\pi\beta
\frac{d^2\tilde{p}_\infty}{d\phi^2}(0)}}.
\end{equation}

Firstly, let us remark that $\overline{P_L(\xi)}=P_L(\overline{\xi})$, 
where the over line means complex conjugation. Let $0<\epsilon<\frac{
e^{\beta\omega/2}-x_\infty}{2}$ be a fixed positive number, and let 
$C_\epsilon$ be a circle centered in $x_\infty$ with radius $\epsilon$. If 
$\xi$ belongs to the interior of $C_\epsilon$ then the Cauchy integral formula 
gives ($L\leq\infty$):
\begin{equation}\label{cortez12}
P_L(\xi)=\frac{1}{2\pi\imath}\int_{C_\epsilon}d\zeta\;
\frac{P_L(\zeta)}{\zeta-\xi}.
\end{equation}

Take $0<\delta<1/2$ and define $\phi_L:=N(L)^{-1/2+\delta/3}$. Then for $L$ 
big enough and $|\phi|\leq \phi_L$, one has that $x_Le^{\imath\phi}$ 
belongs to the interior of $C_{\epsilon/2}$ and (\ref{cortez12}) implies:
\begin{equation}\label{cortez13}
\tilde{p}_L(\phi)=\frac{1}{4\pi\imath}\int_{C_\epsilon}d\zeta\;
P_L(\zeta)\left (\frac{1}{\zeta-x_Le^{\imath\phi}}+
\frac{1}{\zeta-x_Le^{-\imath\phi}}\right ).
\end{equation}

Theorem \ref{ancorc} implies now that:
\begin{equation}\label{cortez14}
 \lim_{L\rightarrow\infty}\frac{d^2\tilde{p_L}}{d\phi^2}(0)=
\frac{d^2\tilde{p_\infty}}{d\phi^2}(0)\quad {\rm and}\quad
 \forall |\phi|\leq\phi_L,\quad\left \arrowvert
 \frac{d^3\tilde{p_L}}{d\phi^3}(\phi)\right\arrowvert\leq {\rm const}.
\end{equation}

Denote with $s_L(\phi)$ the function $\Im P_L(x_Le^{\imath\phi})$. 
In a similar way, one can obtain that for all $|\phi|\leq\phi_L$:
\begin{equation}\label{cortez15}
\left \arrowvert
 \frac{d^3 s_L}{d\phi^3}(\phi)\right\arrowvert\leq {\rm const}.
\end{equation}

Finally, let us prove (\ref{cortez11}). This will be done in two steps: 
\begin{enumerate}
\item The first one consists in showing 
that the contribution to ${\cal A}_L$ coming from the region 
$|\phi|\geq\phi_L$ is exponentially small in $L$; we shall use that 
$\tilde{p}_L(\phi)$ is an even function, is decreasing on the 
interval $[0,\pi]$ and has a non-degenerate maximum in $0$;
\item  The second one 
consists in a more careful study of the case in which $|\phi|\leq\phi_L$; the 
most important thing here is showing 
that the oscillations of the imaginary part are small.
\end{enumerate}

Let us prove 1. Firstly, because $\tilde{p}_L(\phi)=\tilde{p}_L(-\phi)$, 
it is sufficient to look only at $\phi\in [\phi_L,\pi]$. From (\ref{1.2}) one 
has:
\begin{equation}\label{cortez16}
\tilde{p}_L(\phi)=-\frac{1}{2\beta L^3}\sum_j\log
\left [\left (1-x_L\cos\phi\;e^{-\beta E_j}\right )^2+x_L^2\sin^2\phi\;
e^{-2\beta E_j}\right ],
\end{equation}
and:
\begin{equation}\label{cortez17}
\frac{d\tilde{p}_L}{d\phi}(\phi)=-\frac{x_L\sin\phi}{\beta L^3}
\sum_j\frac{e^{-\beta E_j}}{\left\arrowvert 1-x_L e^{\imath\phi}
\;e^{-\beta E_j}\right\arrowvert ^2}\;.
\end{equation}
This means that $\tilde{p}_L$ is decreasing on $[0,\pi]$. 
Its second derivative in $0$ reads as:
\begin{equation}\label{cortez18}
\frac{d^2\tilde{p}_L}{d\phi^2}(0)=-\frac{x_L}{\beta L^3}
\sum_j\frac{e^{-\beta E_j}}{\left\arrowvert 1-x_L
\;e^{-\beta E_j}\right\arrowvert ^2}\;.
\end{equation}
For $L$ big enough, one has that $x_L\geq x_\infty/2$. 
The above equality implies:
\begin{equation}\label{cortez19}
\frac{d^2\tilde{p}_L}{d\phi^2}(0)\leq -\frac{x_\infty}{2\beta}\frac{1}{L^3}
{\rm tr W_L}.
\end{equation}
To obtain an uniform estimate in $L$, we employ (see [K]):
\begin{equation}\label{cortez20}
\frac{1}{L^3}{\rm tr W_L}=\frac{1}{{(2\pi\beta)}^{3/2}}\frac{\omega\beta/2}{
\sinh(\omega\beta/2)}\;(1+{\cal O}(1/L)).
\end{equation}
Then (\ref{cortez14}), (\ref{cortez19}) and (\ref{cortez20}) imply that for 
L sufficiently large:
\begin{equation}\label{cortez21}
\frac{d^2\tilde{p}_L}{d\phi^2}(0)\leq -\frac{x_\infty}{4\beta}
\frac{1}{{(2\pi\beta)}^{3/2}}\frac{\omega\beta/2}{
\sinh(\omega\beta/2)}\quad {\rm and}\quad 
\frac{d^2\tilde{p}_\infty}{d\phi^2}(0)<0.
\end{equation}
From (\ref{cortez21}) and (\ref{cortez14}) it follows that for all $\phi\in
[\phi_L,\pi]$ one has:
\begin{equation}\label{cortez22}
\tilde{p}_L(\phi)-P_L(x_L)\leq \tilde{p}_L(\phi_L)-\tilde{p}_L(0)\leq 
\frac{1}{4}\frac{d^2\tilde{p}_\infty}{d\phi^2}(0)\phi_L^2,
\end{equation}
or in other words,
\begin{equation}\label{cortez23}
\exp{\left \{\frac{N(L)\beta}{\rho} [\tilde{p}_L(\phi)-P_L(x_L)]\right \}}\leq 
\exp{\left [\frac{\beta}{4\rho}\frac{d^2\tilde{p}_\infty}{d\phi^2}(0)
N(L)^{2\delta/3}\right]},
\end{equation}
therefore the contribution to ${\cal A}_L$ coming from $|\phi|\geq\phi_L$ is 
exponentially small.

Now let us study the region $|\phi|\leq\phi_L$. We are mainly interested in 
the behavior of the imaginary part of the pressure (see (\ref{cortez15})). 
Similarly as in (\ref{cortez16}), one has:
\begin{equation}\label{cortez24}
s_L(\phi)=-\frac{1}{\beta L^3}\sum_j\arg{(1-x_Le^{\imath\phi}e^{-\beta E_j})}
=\frac{1}{\beta L^3}\sum_j\arctan{
\frac{x_L\sin\phi\;e^{-\beta E_j}}{1-x_L\cos\phi\;e^{-\beta E_j}}}.
\end{equation}
By direct computation, one obtains:
\begin{equation}\label{cortez25}
s_L(0)=0,\quad \frac{ds_L}{d\phi}(0)=\frac{\rho_L(x_L)}{\beta}=
\frac{\rho}{\beta}\quad {\rm and}\quad 
\frac{d^2s_L}{d\phi^2}(0)=0.
\end{equation}
Together with (\ref{cortez15}), one has:
\begin{equation}\label{cortez26}
\frac{\beta}{\rho}s_L(\phi)-\phi ={\cal O}(\phi_L^3).
\end{equation}
We are interested now in the following integral:
\begin{equation}\label{cortez27}
\frac{\sqrt{N(L)}}{2\pi}\int_{-\phi_L}^{\phi_L} d\phi\;
e^{\left [N(L)\beta/\rho 
(\Re P_L(x_L e^{\imath\phi})-P_L(x_L))\right ]}
e^{\imath N(L)\left [\beta\Im P_L(x_L e^{\imath\phi})/\rho - \phi\right ]}.
\end{equation}
Changing the variable in $t=\sqrt{N(L)}\phi$ one obtains:
\begin{equation}\label{cortez28}
\frac{1}{2\pi}\int_{-N(L)^{\delta/3}}^{N(L)^{\delta/3}}dt\;
e^{\frac{\beta}{2\rho}\frac{d^2\tilde{p}_L}{d\phi^2}(0)\;t^2}
e^{{\cal O}[N(L)^{-1/2+\delta}]}.
\end{equation}
Using (\ref{cortez21}), (\ref{cortez14}) and the 
Lebesgue dominated convergence 
theorem, it is easy to see that the 
above integral converges to (\ref{cortez11}). 

From (\ref{cortez3}) and (\ref{cortez9}), the reduced free energy reads as:
\begin{equation}\label{cortez29}
f_L=\tilde{f}_L+\frac{\log N(L)}{2\beta L^3}-\frac{\log {\cal A}_L}{\beta L^3}
\; .
\end{equation}
From (\ref{cortez4}) and (\ref{cortez5}) it follows:
\begin{equation}\label{cortez30}
m_L=\tilde{m}_L-\frac{e}{\beta c L^3}\;\frac{1}{{\cal A}_L}\;
\frac{\partial {\cal A}_L}{\partial\omega},
\end{equation}
or:
\begin{eqnarray}\label{cortez31}
&{}& m_L-\tilde{m}_L = \\
&=& \frac{N(L)^{1/2}}{2\pi\;{\cal A}_L}\int_{-\pi}^{\pi}d\phi\;
e^{\left [\frac{N(L)\beta}{\rho}\left (P_L(x_Le^{\imath\phi})-P_L(x_L)\right )
\right ]}\left [-\frac{e}{c}\frac{d}{d\omega}\left (P_L(x_Le^{\imath\phi})-
P_L(x_L)\right )\right ],\nonumber
\end{eqnarray}
where:
\begin{eqnarray}\label{cortez32}
&{}& -\frac{e}{c}\frac{d}{d\omega}\left (P_L(x_Le^{\imath\phi})-
P_L(x_L)\right )= \\ 
&=& \Gamma_L(x_Le^{\imath\phi})-\Gamma_L(x_L)-\frac{e}{c}
\left [\frac{\partial P_L}{\partial z}(x_Le^{\imath\phi})
\frac{\partial x_L}{\partial \omega}e^{\imath\phi}-
\frac{\partial P_L}{\partial z}(x_L)
\frac{\partial x_L}{\partial \omega}\right ].\nonumber
\end{eqnarray}

From (\ref{1.5}) it follows that 
$-\frac{e}{c}\frac{\partial\rho_L}{\partial\omega}(\beta,z,\omega)=\beta z
\frac{\partial \Gamma_L}{\partial z}(\beta,z,\omega)$. This implies:
$$-\frac{e}{c}\;\frac{\partial x_L}{\partial \omega}=
-\beta x_L \frac{\partial \Gamma_L}{\partial z}(x_L)
\left [\frac{\partial \rho_L}{\partial z}(x_L)\right ]^{-1}.$$
Lemma \ref{limgama} and Theorem \ref{ancorc} imply (via the Cauchy  integral 
formula) that the above quantity remains bounded when $L$ goes to infinity. 
Performing a similar analysis of the integral from (\ref{cortez31}) as 
that one made for ${\cal A}_L$, it follows that the contribution coming from 
the region $|\phi|\geq \phi_L$ is exponentially small in $L$. When we are 
analyzing the region $|\phi|\leq\phi_L$, the factor $N(L)^{1/2}$ disappears 
when we are changing the variable, therefore $m_L-\tilde{m}_L$ will behave as 
the quantity from (\ref{cortez32}), and this one behaves like $\phi_L\sim 
L^{-3/2+\delta}$. The proof of (\ref{cortez6}) is now completed.
\section{The Proof of Lemma \ref{limgama}}
\setcounter{equation}{0}

The idea of the proof is borrowed from [A]. Namely, the uniform 
convergence on the compacts which belong to ${\cal D}:={\CM}\setminus 
[e^{\beta\omega_0/2},\infty)$ 
will be obtained applying the Vitali  Theorem to 
the sequence of analytical functions $\Gamma_L(\beta,z,\omega_0)$ when $L$ is 
going to infinity. Therefore, we have to make the following three steps:

I. Prove the uniform boundedness in $L$ of the functions 
$\Gamma_L(\beta,z,\omega_0)$ on any compact $K\subset {\cal D}$;

II. Identify the limit ($\Gamma_\infty(\beta,z,\omega_0)$ in our case) 
and prove that it has the same domain of 
analyticity ${\cal D}$;

III. Prove the existence of a set ${\cal D}_0\subset {\cal D}$ having at 
least one point of accumulation, such that $\Gamma_L(\beta,z,\omega_0)$ has 
a pointwise convergence to $\Gamma_\infty(\beta,z,\omega_0)$ on ${\cal D}_0$.

\subsection{The proof of I}

Here is the ``nontrivial part'' of our paper. In order to see where the main 
difficulty is, let us take a look at (\ref{usd'}). One could try a bound on 
$\Gamma_L(\beta,z,\omega)$ using the following inequality (see also 
(\ref{dewlaom})):
\begin{eqnarray}
\sup_{z\in K}\left\arrowvert {\rm tr}\left [(1-zW_L)^{-1}\frac{\partial W_L}
{\partial\omega}\right ]\right\arrowvert &\leq& 
\left\Arrowvert\frac{\partial W_L}{\partial\omega}\right\Arrowvert_
{{\cal I}_1}
\;\sup_{z\in K}\left\Arrowvert (1-zW_L)^{-1}
\right\Arrowvert_
{B({\bf L}^2)}\nonumber \\
&\leq&C(\beta,K,\omega)
\left\Arrowvert\frac{\partial W_L}{\partial\omega}\right\Arrowvert_
{{\cal I}_1}.\nonumber
\end{eqnarray}

Unfortunately, the linear growth of the vector potential 
${\bf a}$ leads to a bad trace norm 
estimate for $\frac{\partial W_L}{\partial\omega}$, which behaves like $L^4$ 
and not like $L^3$ as needed. In the case of the pressure [A-C], this kind of 
problem does not appear (see (\ref{1.15}), (\ref{1.14}) and (\ref{1.12})), 
because for the Gibbs semigroup the trace and the trace norm are equal and 
grow like $L^3$. 

Because the proof is rather lengthy, we shall outline in a few words 
our strategy. Firstly, having (\ref{1.14}) in mind, let us define 
($\omega_0,\beta,\tau >0$):
\begin{equation}\label{gbetatau}
g_{\omega_0}(\xi,z;\beta,\tau):=
[\xi-zW_L(\beta,\omega_0)]^{-1}zW_L(\tau,\omega_0),
\end{equation}
where $\xi$ belongs to the contour ${\cal C}$. 
Then the pressure will admit the following representation (see (\ref{1.15})):
\begin{equation}\label{reppresi}
P_L(\beta,z,\omega_0)=-\frac{1}{2\pi\imath}
\int_{\cal C}d\xi\;q(\xi)\frac{1}{L^3}\;
{\rm tr}[g_{\omega_0}(\xi,z;\beta,\beta)],
\end{equation}
where the norm convergent integral from (\ref{1.14}) commutes 
with the trace from (\ref{1.15}). Let us argue that we can choose the 
same contour of integration if $\omega$ varies in a small interval 
$\Omega:=[\omega_0,\omega_1]$, $0<\omega_0<\omega_1$. In other words, we 
search for a 
${\cal C}$ included in ${\CM}\setminus [1,\infty )$ such that for all $L>1$, 
$z\in K$, $\omega\in\Omega$ and $\xi\in {\cal C}$ one has the following bound:
\begin{equation}\label{K1} 
\Arrowvert [\xi -zW_L(\beta,\omega)]^{-1}\Arrowvert \leq M < \infty.
\end{equation}
Assume that every $z\in K$ verifies the following condition:
\begin{equation}\label{K} 
 {\rm dist}\{ z, \left [ \right. e^{\frac{\beta\omega_0}{2}},\infty 
\left. \right )  
 \}\geq\delta >0, \quad |z|\leq d<\infty. 
\end{equation}
We know that the spectrum of $W_L(\beta,\omega)$ is included in 
$[0, e^{-\frac{\beta\omega_0}{2}}]$ 
for all $L>1$ and $\omega\in\Omega$.  We claim that ${\cal C}$ can be chosen 
as the union 
${\cal C}_1\cup {\cal C}_2$, where ${\cal C}_2$ is given by ($\eta >0$):
$$\{(1 +t,\pm \eta)|\;-\eta\leq t\leq 2d \}\cup
\{(1-\eta,t)|-\eta\leq t\leq\eta\},$$
and ${\cal C}_1$ is chosen such that if $\xi\in {\cal C}_1$, then 
$|\xi|\geq 2d +1$.
It is not difficult to prove that by choosing $\eta $ sufficiently small, 
then: 
\begin{equation}\label{K2}
\sup_{z\in K}\sup_{\xi\in {\cal C}}
\sup_{0\leq r\leq e^{-\frac{\beta\omega_0}{2}}}|(\xi -z\:r)^{-1}|\leq M 
<\infty,
\end{equation}
and via the Spectral Theorem, (\ref{K1}) takes place.

If we manage to prove the existence of a numerical constant 
$c(\beta,K,\omega_0)$ such that for all 
$\delta\omega\in (0,\omega_1-\omega_0)$ and $z\in K$ to have:
\begin{equation}\label{sgda}
\sup_{z}\sup_{\delta\omega}
\frac{1}{\delta\omega}|P_L(\beta,z,\omega_0+\delta\omega)-
P_L(\beta,z,\omega_0)|\leq c(\beta,K,\omega_0),
\end{equation}
then the magnetization will be bounded by the same constant. 
This estimate is straightforward if a stronger one takes place: 
\begin{equation}\label{sgda'}
\sup_{z}\sup_\xi\sup_{\delta\omega}
\frac{1}{L^3\delta\omega}
|{\rm tr}[g_{\omega_0+\delta\omega}(\xi,z;\beta,\beta)]-
{\rm tr}[g_{\omega_0}(\xi,z;\beta,\beta)]|\leq C(\beta,K,\omega_0).
\end{equation}
Our main task will consist in constructing a trace class operator 
$A_{\omega_0+\delta\omega}(\xi,z;\beta)$ having the following two properties: 
${\rm tr}A_{\omega_0+\delta\omega}(\xi,z;\beta)=
{\rm tr}g_{\omega_0}(\xi,z;\beta,\beta)$ (i.e. its trace is {\it not} 
depending on $\delta\omega$) and moreover,
\begin{equation}\label{sgda''} 
\sup_{z}\sup_\xi\sup_{\delta\omega}
\frac{1}{L^3\delta\omega}
\Arrowvert g_{\omega_0+\delta\omega}(\xi,z;\beta,\beta)-
A_{\omega_0+\delta\omega}(\xi,z;\beta)\Arrowvert_{{\cal I}_1}\leq C'(\beta,K,
\omega_0),
\end{equation}
which would clearly end the problem. 

We will see that ($\omega=\omega_0+\delta\omega$) 
$A_{\omega}(\xi,z;\beta)$ can be chosen as a product 
${\tilde g}_{\omega}(\xi,z;\beta,\beta/2)
S_{\omega}(\beta/2)$ where the first term is bounded, the 
second one is trace class and moreover:
\begin{equation}\label{heurtu}
\sup_{z}\sup_\xi\Arrowvert g_{\omega}(\xi,z;\beta,\beta/2)-
{\tilde g}_{\omega}(\xi,z;\beta,\beta/2 )
\Arrowvert_{B({\bf L}^2)}\leq C_1(\beta,K,\omega_0)\delta\omega ,
\end{equation}

and

\begin{equation}\label{heurtu'}
\Arrowvert W_L(\beta/2 ,\omega)-
S_{\omega}(\beta/2)\Arrowvert_{{\cal I}_1}\leq 
 C_2(\beta,\omega_0)L^3\delta\omega.
\end{equation}  
In particular, (\ref{heurtu}) and (\ref{heurtu'}) imply:
\begin{equation}
\Arrowvert 
S_{\omega}(\beta/2)
\Arrowvert_{{\cal I}_1}\leq C(\beta,\omega_0)L^3\; {\rm and}\; 
\sup_{z}\sup_\xi\Arrowvert  
{\tilde g}_{\omega}(\xi,z;\beta,\beta/2 ) 
\Arrowvert_{B({\bf L}^2)}\leq C_3(\beta,K,\omega_0).
\end{equation}
Employing these estimates together with 
the following identity (see (\ref{gbetatau})): 
$$g_{\omega}(\xi,z;\beta,\beta)=
g_{\omega}(\xi,z;\beta,\beta/2)
W_L(\beta/2 ,\omega),$$ the proof of (\ref{sgda''}) follows 
easily.

The rest of this subsection is dedicated to the rigorous proofs of these 
estimates and will be structured in a sequence of technical propositions. 
We start with a well known result, given without any other comments:
\begin{proposition}\label{p1}
The Dirichlet Laplacian defined in $\Lambda$ admits a trace class semigroup 
which has an integral kernel given by the following formula:
\begin{equation}\label{prl1}
G_{0,L}({\bf x},{\bf x'};\beta)=\prod_{j=1}^3 g_{0,L}(x_j,x_j';\beta),
\end{equation}
where the ``one dimensional'' kernels read as:
\begin{eqnarray}\label{omicron}
&{}& g_{0,L}(x,x';\beta)=\nonumber \\ 
&=& \frac{1}{(2\pi\beta)^{1/2}}
\sum_{m\in {\bf Z}}\left \{
\exp{\left [-\frac{(x-x'+2mL)^2}{2\beta}\right ]}-
\exp{\left [-\frac{(x+x'-2mL-L)^2}{2\beta}\right ]}
\right \} \nonumber \\
&:=& g_{0,\infty}(x,x';\beta)+\zeta_{0,L}(x,x';\beta).
\end{eqnarray}
\end{proposition}
\vspace{0.5cm}

Using the previous proposition, one can write:
\begin{equation}
G_{0,L}({\bf x},{\bf x'};\beta)=
G_{0,\infty}({\bf x},{\bf x'};\beta)+Z_{0,L}({\bf x},{\bf x'};\beta),
\end{equation}
where:
\begin{equation}\label{liberta}
G_{0,\infty}({\bf x},{\bf x'};\beta)=
\frac{1}{(2\pi\beta)^{3/2}}\exp{\left ( -\frac{|{\bf x}-{\bf x'}|^2}
{2\beta}\right)}.
\end{equation}

The purpose of the next proposition is to give a few properties of 
smoothness and localization of the reminder $Z_{0,L}({\bf x},{\bf x'};\beta)$:
\begin{proposition}\label{p2}
For all $\beta >0$ and $L>1$, 
there exist two positive numerical constants $c_1$ and $c_2$ such that:
\begin{eqnarray}\label{unu}
{\rm i.} &{}&
\left\arrowvert \frac{\partial Z_{0,L}}{\partial x_j}\right\arrowvert
({\bf x},{\bf x'};\beta) \leq
 c_1 \:\frac{1+\beta}{\beta^{1/2}}\: 
G_{0,\infty}({\bf x},{\bf x'};c_2\beta);\nonumber \\
{\rm ii.} &{}& 
\max {\left\{
\left\arrowvert \frac{\partial Z_{0,L}}{\partial \beta}\right\arrowvert
({\bf x},{\bf x'};\beta),
\left\arrowvert \frac{\partial^2 Z_{0,L}}{\partial x_j\partial x_k'}
\right\arrowvert({\bf x},{\bf x'};\beta)
\right \} }\leq  c_1\frac{1+\beta}{\beta} 
G_{0,\infty}({\bf x},{\bf x'};c_2\beta);\nonumber \\
{\rm iii.} &{}&
\arrowvert Z_{0,L}\arrowvert({\bf x},{\bf x'};\beta)\leq
 c_1 \:(1+\beta)\: 
G_{0,\infty}({\bf x},{\bf x'};c_2\beta).\nonumber
\end{eqnarray}
\end{proposition}

{\it Proof.} For $x,x'\in (-L/2,L/2)$ define:
\begin{eqnarray}\label{xi1,2}
\zeta_1(x,x';\beta) &=& \frac{1}{\sqrt{2\pi\beta}}
\sum_{m\in {\bf Z}\setminus \{0\}}\exp{\left (-\frac{(x-x'+2mL)^2}
{2\beta}\right )},\; \\
\zeta_2(x,x';\beta) &=& \frac{1}{\sqrt{2\pi\beta}} 
\sum_{m\in {\bf Z}}\exp{\left (-\frac{[x-x'-(2m+1)L]^2}
{2\beta}\right )}.\nonumber
\end{eqnarray}

It is clear that if one obtains uniform estimates in $L$ and $\beta$ for 
these two quantities, the same would be true for $Z_{0,L}$, too. A very useful 
estimate is the following:
\begin{equation}\label{elat}
\forall t\geq 0,\quad t e^{-t}=2(t/2)e^{-t/2}e^{-t/2}\leq 2e^{-t/2}.
\end{equation}
Then we have the inequalities:
\begin{eqnarray}\label{xi'}
&{}& \left\arrowvert\frac{\partial\zeta_1}{\partial x}\right\arrowvert
(x,x';\beta) \leq\frac{{\rm const.}}{\beta}
\sum_{m\neq 0}\exp{\left (-\frac{(x-x'+2mL)^2}
{4\beta}\right )},  \\
&{}&\left\arrowvert\frac{\partial^2\zeta_1}{\partial x\partial x'}
\right\arrowvert (x,x';\beta)
\leq  
 \frac{{\rm const.}}{\beta^{3/2}}
\sum_{m\neq 0}\exp{\left (-\frac{(x-x'+2mL)^2}
{4\beta}\right )},\nonumber 
\end{eqnarray}
and similarly for $\zeta_2$. At this point we have to control the summation 
over $m$. Let us prove the following inequality:
\begin{equation}\label{giugude}
\sum_{m\neq 0}\exp{\left (-\frac{(x-x'+2mL)^2}
{4\beta}\right )}\leq c_1(1+\beta)\exp{\left (-\frac{(x-x')^2}
{c_2\beta}\right )}.
\end{equation}
Because $|x-x'|< L$ one has ($|m|,L\geq 1$):
\begin{equation}\label{xsiL}
(x-x'+2mL)^2=(x-x')^2+4mL(x-x')+4m^2L^2\geq (x-x')^2+4(|m|-1).
\end{equation}
Therefore:
\begin{eqnarray}\label{castilio'}
&{}& \sum_{m\neq 0}\exp{\left (-\frac{(x-x'+2mL)^2}
{4\beta}\right )}\leq \nonumber \\
&\leq & \exp{\left (-\frac{(x-x')^2}
{4\beta}\right )}2\left [1+\sum_{m\geq 1}\exp{\left (-\frac{1}
{\beta}m\right )}\right ]=\nonumber \\
&=& 2\exp{\left (-\frac{(x-x')^2}
{4\beta}\right )}\left [1+\frac{1}{2}\frac{\frac{1}
{2\beta}}{\sinh{\left (\frac{1}{2\beta}\right )}}
2\beta\exp{\left (-\frac{1}
{2\beta}\right )}\right ]\leq \nonumber \\
&\leq & 2(1+\beta)\exp{\left (-\frac{(x-x')^2}
{4\beta}\right )} .
\end{eqnarray}
In order to control $\zeta_2$, one has to study the following quantity:
\begin{equation}\label{1432}
A:=\sum_{m\in {\bf Z}}\exp{\left (-\frac{[x+x'-(2m+1)L]^2}
{4\beta}\right )}.
\end{equation}
Denote with $\xi=x+L/2$ and $\xi'=x'+L/2$; then $0<\xi,\xi'<L$, 
$x-x'=\xi-\xi'$ and $(\xi+\xi')^2\geq (\xi-\xi')^2$. It follows:
\begin{eqnarray}\label{14323}
[x+x'-(2m+1)L]^2 &=&[\xi+\xi'-2(m+1)L]^2 \\
&=& (\xi+\xi')^2-4(m+1)L(\xi+\xi')+
4(m+1)^2L^2. \nonumber
\end{eqnarray}
If $m\leq -1$, then:
$$[x+x'-(2m+1)L]^2\geq (x-x')^2+4(|m|-1).$$
If $m=0$, then:
$$[x+x'-(2m+1)L]^2=[(L/2-x)+(L/2-x')]^2\geq (x-x')^2.$$
If $m\geq 1$, then:
$$[x+x'-(2m+1)L]^2\geq (\xi+\xi')^2+4L^2(m+1)(m-1)\geq (x-x')^2+4(m-1),$$
and we can repeat the summation procedure used in (\ref{giugude}). Putting 
all these things together, the proof is completed.

\vspace{0.5cm}

The next proposition is a variant of the perturbation theory for self-adjoint 
Gibbs semigroups (see [H-P]). Instead of starting with a perturbation of its 
generator, we start with an approximation of the semigroup. Although simple, 
this proposition contains the main technical core of our paper.
\begin{proposition}\label{general1}
Let  ${\cal H}:={\bf L}^2(\Lambda)$ and let $H$ be a self-adjoint and positive 
operator having the domain $D$. Fix $\beta_0>0$. Assume that there exists an 
application $0<\beta \leq\beta_0\rightarrow S(\beta)\in B({\cal H})$ with 
the following properties:

{\rm A}. $\sup_{0<\beta\leq\beta_0}||S(\beta)||\leq c_1<\infty$;

{\rm B}. It is strongly differentiable, ${\rm Ran}S(\beta)\subset D$ and 
$s-\lim_{\beta\searrow 0}S(\beta)=1$;

{\rm C}. There exists a normly continuous application 
$0<\beta \leq\beta_0\rightarrow R(\beta)\in B({\cal H})$ such that 
$||R(\beta)||\leq c_2/ \beta^{\alpha}$ where $0\leq\alpha <1$ and:
\begin{equation}\label{duclos}
\frac{\partial S}{\partial\beta}f+HS(\beta)f=R(\beta)f.
\end{equation}

Then the following two statements are true:

{\rm i}. The sequence of bounded operators ($n >[1/\beta]$): 
$$ T_n(\beta):=\int_{1/n}^{\beta -1/n}d\tau\; \exp{[-(\beta -\tau)H]}R(\tau)$$ 
converges in norm; let $T(\beta)$ be its limit;

{\rm ii}. The following equality takes place in $B({\cal H})$:
\begin{equation}\label{Briet}
\exp{(-\beta H)}=S(\beta)-T(\beta).
\end{equation}
\end{proposition}
{\it Proof.} i. The norm convergence is assured by the integrability 
condition imposed on the norm of $R(\beta)$. Moreover, when $\beta$ is near 
zero:
\begin{equation}\label{Brietp}  
\sup_n ||T_n(\beta)||\leq c(\alpha)\beta^{1-\alpha},
\end{equation}
therefore the same thing is true for $T(\beta)$.

ii. Let  $0<\beta_1 <\beta <\beta_0$. If  $n> [1/\beta_1]$ and 
$\phi\in {\cal H}$, define the vector:
\begin{equation}\label{BrietP}
\psi_n(\beta):=\exp{(-\beta H)}\phi -S(\beta)\phi +T_n(\beta)\phi.
\end{equation}
From (\ref{Brietp}) and condition A it follows:
\begin{equation}\label{psin}
\lim \psi_n(\beta) := 
\psi(\beta)=\exp{(-\beta H)}\phi -S(\beta)\phi +T(\beta)\phi\;
 {\rm and}\;
\sup_{n,\beta}||\psi_n(\beta)|| \leq  {\rm const}.
\end{equation}
Define $f_n(\beta)=||\psi_n(\beta)||^2$ and $f(\beta)=||\psi(\beta)||^2$. 
From the strong convergence to one of  $S(\beta)$ when $\beta$ goes to zero 
and from the norm convergence 
to zero of $T(\beta)$, it follows that  $\lim_{\beta\searrow 0}f(\beta)=0$. 
If we manage to prove that $f(\beta)$ is decreasing, then it would be 
identically zero and this would end the proof.

Notice that $T_n(\beta)$ is normly differentiable and:
\begin{equation}\label{Combes}
\frac{\partial T_n}{\partial\beta}\phi +H T_n(\beta)\phi =
\exp{\left (-\frac{1}{n} H \right )}
R(\beta -1/n)\phi .
\end{equation}
The positivity of $H$ implies:
\begin{equation}\label{Combes1}
\frac{\partial f_n}{\partial\beta}\leq 2 \;\Re
[\langle\psi_n(\beta),\exp{\left (-\frac{1}{n} H \right )}
R(\beta -1/n)\phi -R(\beta)\phi\rangle ].
\end{equation}
Using $f_n(\beta)=f_n(\beta_1)+\int_{\beta_1}^\beta d\tau\;
\frac{\partial f_n}{\partial\tau}$, one has:
\begin{equation}\label{Combes2}
0\leq f_n(\beta)\leq f_n(\beta_1)+{\rm const}\int_{\beta_1}^\beta d\tau\;
||\exp{\left (-\frac{1}{n} H \right )}
R(\tau -1/n)\phi -R(\tau)\phi||.
\end{equation}
Because the above integrand is bounded on the domain of integration and 
has pointwise convergence to zero, the dominated convergence theorem implies 
that the whole integral converges to zero. Taking the limit, it 
follows that $f(\beta)$ is decreasing and therefore is identically zero.

\vspace{0.5cm}

Define the ``magnetic phase'':
\begin{equation}\label{faza}
\varphi({\bf x},{\bf x'})={\bf x}\cdot{\bf a}( {\bf x'})=\frac{1}{2}
{\bf e_3}\cdot ({\bf x'}\wedge {\bf x}).
\end{equation}
We shall use this quantity as a local gauge transformation (see [C-N] for 
another use of this idea); namely, it will alter the 
magnetic vector potential by making it depend on ${\bf x}-{\bf x'}$, only: 
${\bf a}({\bf x})-{\bf a}({\bf x}-{\bf x'})=
\nabla_{{\bf x}}\varphi({\bf x},{\bf x'})$. 
To see how this transformation acts (at a formal 
level) on $H_{1,L}(\omega)$, let us notice the following equation:
\begin{equation}\label{fazada}
\left [-\imath\nabla_{{\bf x}}-\omega{\bf a}({\bf x})\right ]
e^{\imath\omega\varphi({\bf x},{\bf x'})}=
e^{\imath\omega\varphi({\bf x},{\bf x'})}
\left [-\imath\nabla_{{\bf x}}-\omega{\bf a}({\bf x}-{\bf x'})\right ].
\end{equation}
\begin{proposition}\label{pert}
The bounded operator $S(\beta)\in B\left ({\bf L}^2 (\Lambda)\right )$ 
given by the 
integral kernel $e^{\imath\omega\varphi({\bf x},{\bf x'})}
G_{0,L}({\bf x},{\bf x'};\beta)$, verifies the hypotheses of Proposition 
\ref{general1}, having the following properties:

{\rm i}. ${\rm s}-\lim_{\beta\searrow 0}S(\beta)={\bf 1}$;

{\rm ii}. The application 
$(0,\infty)\ni \beta \longmapsto S(\beta)$ is strongly differentiable 
and uniformly bounded;

{\rm iii}. ${\rm Ran}S(\beta)\in {\rm Dom}(H_{1,L}(\omega))$ 
and 
$\frac{\partial S}{\partial\beta}(\beta)f+
H_{1,L}(\omega)S(\beta)f=R(\beta)f$, 
where $R(\beta)$ has an integral kernel 
$R({\bf x},{\bf x'};\beta)$ given by:
$$
 e^{\imath\omega\varphi({\bf x},{\bf x'})}
[\omega^2{\bf a}^2({\bf x}-{\bf x'})G_{0,L}({\bf x},{\bf x'};\beta)
+2\imath \omega {\bf a}({\bf x}-{\bf x'})\cdot \nabla_{\bf x} 
Z_{0,L}({\bf x},{\bf x'};\beta)].$$
\end{proposition}
{\it Proof.} i. Throughout the whole section, we shall use a few well known 
results, which are given without proof. Let us begin with an useful 
boundedness criterion 
for integral operators (see [S]):

Let $A$ be an integral operator, given by a continuous integral kernel 
$A({\bf x},{\bf x'})\in
C(\overline{\Lambda}\times\overline{\Lambda})$. If the next inequality 
holds:
\begin{equation}\label{9999}
\max\left [\sup_{{\bf x}\in\Lambda}\int_{\Lambda}d{\bf x'}\;
|A({\bf x},{\bf x'})|,
\sup_{{\bf x'}\in\Lambda}\int_{\Lambda}d{\bf x}\;
|A({\bf x},{\bf x'})|\right ]\leq {\rm C}<\infty,
\end{equation}
then the operator norm of $A$ in $B({\bf L}^2(\Lambda))$ is bounded by 
${\rm C}.$

If $c$, $\beta$ and $\tau <\beta$ are three strictly positive numbers, then 
we have the following three identities:
\begin{equation}\label{9999'}
\int_{{\RM}^3}d{\bf y}\;
\exp{\left (-\frac{|{\bf x}-{\bf y}|^2}{c(\beta-\tau)}-
\frac{|{\bf y}-{\bf x'}|^2}{c\tau}\right )}=
 (\pi c)^{\frac{3}{2}}\left [\frac{(\beta -\tau)\tau}{\beta}\right ]^
{\frac{3}{2}}\exp{\left (-\frac{|{\bf x}-{\bf x'}|^2}{c\beta}\right )},
\end{equation}
\begin{equation}\label{9999''}
\int_{{\RM}^3}d{\bf y}\;
\frac{1}{(\pi c\beta)^{\frac{3}{2}}}
\exp{\left (-\frac{|{\bf x}-{\bf y}|^2}{c\beta}\right )}=1,
\end{equation}
\begin{equation}\label{9999'''}
\left (\int_{{\RM}^3}d{\bf y}\;|G_{0,\infty}({\bf x},{\bf y};\beta)|^2
\right )^{\frac{1}{2}}=(2\pi\beta)^{-\frac{3}{4}}.
\end{equation}
A very useful inequality will be the next one ($t>0,\;n\geq 1$):
\begin{equation}\label{ineh}
t^n\exp{\left (-\frac{t^2}{\beta}\right )} \leq 
{\rm const}(n)\beta^{\frac{n}{2}}\exp{\left (-\frac{t^2}{2\beta}\right )}.
\end{equation}

Let us get back to the proof of i. Firstly, because 
$|e^{\imath\omega\varphi({\bf x},{\bf x'})}
G_{0,L}({\bf x},{\bf x'};\beta)|\leq G_{0,\infty}({\bf x},{\bf x'};\beta)$, 
it follows that $S(\beta)$ obeys the condition (\ref{9999}) with 
${\rm C}\leq 1$, which means that is uniformly bounded in $\beta >0$. 

Let us show now that the operator given by the integral kernel 
$(e^{\imath\omega \varphi 
({\bf x},{\bf x'})}-1)G_{0,L}({\bf x},{\bf x'};\beta)$ converges in norm to 
zero. Let us remark first that $|\varphi({\bf x},{\bf x'})|\leq L\;|{\bf x}-
{\bf x'}|$ and moreover:
\begin{equation}\label{strly}
|(e^{\imath\omega \varphi 
({\bf x},{\bf x'})}-1)|\leq \omega |\varphi({\bf x},{\bf x'})|\leq \omega
 L\;|{\bf x}-{\bf x'}|.
\end{equation}
Then (using (\ref{ineh}) with $n=1$):
\begin{equation}\label{strlyy}
|(e^{\imath\omega \varphi 
({\bf x},{\bf x'})}-1)G_{0,L}({\bf x},{\bf x'};\beta)|\leq 
{\rm const}\;L\;\beta^{1/2}G_{0,\infty}({\bf x},{\bf x'};2\beta),
\end{equation}
therefore its operator norm behaves in zero like $\beta^{1/2}$ at least. 
The proof of i is now straightforward. 

ii. We will prove that the application is in fact normly differentiable. For 
$\beta >0$ and $\delta\beta$ sufficiently small, one has:
\begin{eqnarray}\label{stryu}
S(\beta+\delta\beta)f-S(\beta)f &=&
\delta\beta\int_\Lambda d{\bf x'}\;e^{\imath\omega \varphi 
(\cdot,{\bf x'})}
\frac{\partial G_{0,L}}
{\partial\beta}(\cdot,
{\bf x'};\beta)f({\bf x'})+ \nonumber \\
&+& \frac{(\delta\beta)^2}{2}
\int_\Lambda d{\bf x'}\;e^{\imath\omega \varphi 
(\cdot,{\bf x'})}
\frac{\partial^2 G_{0,L}}{\partial\beta^2}(\cdot,
{\bf x'};\tilde\beta)f({\bf x'}),
\end{eqnarray}
where $\tilde\beta$ is situated between  $\beta$ and $\beta +\delta\beta$. 
It is not difficult now to see that the ``operator derivative'' is an 
integral operator whose kernel is the derivative with respect to 
$\beta$ of the initial one.

iii. Denote with $D_0$ the common domain of essentially self-adjointness for 
$H_{1,L}(\omega)$, $\omega\geq 0$:
\begin{equation}\label{oreir}
D_0=\{\psi\in C^2(\Lambda)\cap C^1(\overline{\Lambda})|
\;\psi|_{\partial\Lambda}
=0,\;\Delta\psi\in {\bf L}^2(\Lambda)\}.
\end{equation}
The action of $H_{1,L}(\omega)$ on a function from $D_0$ is as follows:
\begin{equation}\label{oreir1}
[H_{1,L}(\omega)\psi]({\bf x})=-(\Delta\psi)({\bf x})+2\imath\omega
 {\bf a}({\bf x})\cdot (\nabla\psi)({\bf x})+\omega^2 {\bf a}^2({\bf x})\psi(
{\bf x}).
\end{equation}
Now take $f\in C_0^\infty(\Lambda)$. After integration by parts, using 
(\ref{fazada}) and the fact that $G_{0,L}({\bf x},{\bf x'};\beta)$ solves 
the heat equation in the interior of $\Lambda$, one obtains:
\begin{eqnarray}\label{oreir2}
&{}&\langle H_{1,L}(\omega)\psi,S(\beta)f\rangle =
\int_{\Lambda^2} d{\bf x'}\; d{\bf x}\;
\overline{\psi({\bf x})}f({\bf x'})e^{\imath\omega \varphi 
({\bf x},{\bf x'})}
\cdot \nonumber \\
&\cdot& [-\Delta_{{\bf x}}+2\imath\omega
 {\bf a}({\bf x}-{\bf x'})\cdot\nabla_{{\bf x}}+
\omega^2 {\bf a}^2({\bf x}-{\bf x'})]G_{0,L}({\bf x},{\bf x'};\beta)
\nonumber \\ 
&=& -\langle\psi, S'(\beta)f\rangle +\langle\psi, R(\beta)f\rangle.
\end{eqnarray}
The result follows easily after a density argument and with the remark 
that:
\begin{equation}\label{oreir10}
 {\bf a}({\bf x}-{\bf x'})\cdot\nabla_{{\bf x}}
G_{0,\infty}({\bf x},{\bf x'};\beta)=0.
\end{equation}

Finally, let us notice that the norm of $R(\beta)$ is independent of $L$ and 
is integrable in zero. To do that, one has to employ the estimates from 
Proposition \ref{p2}, (\ref{ineh}) and the criterion from (\ref{9999}).

\vspace{0.5cm}

In order to perform a similar 
perturbative treatment of the semigroup near a nonzero magnetic field, we need 
the estimate given by the next proposition:
\begin{proposition}
Let ${\bf n}$ be a unit vector in ${\RM}^3$. Then there exist three positive 
numerical constants $s$, $c_4$ and $c_5$ such that for all $\omega \in\Omega$, 
${\bf x},{\bf x'}\in \Lambda $ and $\beta >0$, one has the following uniform 
estimate in $L$:
\begin{equation}\label{sase}
|{\bf n}\cdot (-\imath \nabla -\omega {\bf a}({\bf x}))
G_{\omega,L}({\bf x},{\bf x'};\beta)|\leq c_4\:\frac{(1+\beta)^s}{\beta^2}\:
\exp{\left [  
-\frac{|{\bf x}- {\bf x'}|^2}{c_5 \beta}
\right ]}.
\end{equation}
\end{proposition}
{\it Proof.} Proposition \ref{general1} allows us to write down the following 
integral equation:
\begin{equation}\label{Hillep}
W_L(\beta,\omega)=S(\beta)-\int_0^\beta d\tau\;
W_L(\beta-\tau,\omega)\;R(\tau).
\end{equation}
Because $S(\beta)$ and $W_L$ are self-adjoint, one can rewrite (\ref{Hillep}) 
as:
\begin{equation}\label{Hillep'}
W_L(\beta,\omega)=S(\beta)-\int_0^\beta d\tau\;R^\ast (\tau)
W_L(\beta-\tau,\omega).
\end{equation}
In terms of integral kernels, (\ref{Hillep'}) reads as:
\begin{equation}\label{Hillep''}
G_{\omega,L}({\bf x},{\bf x'} ;\beta) = 
S({\bf x} ,{\bf x'} ;\beta)-
\int_0^\beta d\tau\int_{\Lambda}d{\bf y}\; 
R^\ast ({\bf x},{\bf y} ,\tau)G_{\omega,L}({\bf y},{\bf x'}, ;\beta -\tau),
\end{equation}
where the equality is between continuous functions in 
$C(\Lambda\times\Lambda)$ and the integral in $\tau$ has to be understood as 
``$\int_\epsilon^{\beta-\epsilon}$'' in the limit $\epsilon\searrow 0$. 

Because the kernel of $R^\ast$ reads as 
$R^\ast ({\bf x},{\bf y};\tau)=\overline{R({\bf y},{\bf x};\tau)}$, by direct 
computation one can obtain the estimate (see Proposition \ref{p2}):
\begin{equation}\label{Rast'}
|{\bf n}\cdot (-\imath \nabla -\omega {\bf a}({\bf x}))
R^\ast({\bf x},{\bf x'};\tau)|\leq c_4'\;\frac{(1+\tau)^{s'}}{\tau^2}
\exp{\left [-\frac{|{\bf x}- {\bf x'}|^2}{c_5' \tau}
\right ]},
\end{equation}
where one has to apply (\ref{fazada}) then use the estimates from 
Proposition \ref{p2}; finally, introducing (\ref{Rast'}), 
(\ref{1.11}), (\ref{9999'}) and 
(\ref{ineh}) in (\ref{Hillep''}) and because the singularity in 
$\tau$ is integrable, the result for $G_{\omega,L}$ is straightforward.

\vspace{0.5cm}

Take $\omega=\omega_0+\delta\omega\in\Omega$. The analogous of Proposition 
\ref{pert} at nonzero magnetic field is:
\begin{proposition}\label{pertb}
The bounded operator denoted with $S_{\omega}(\beta)$ and given by the 
kernel $e^{\imath\delta\omega\varphi({\bf x},{\bf x'})}
G_{\omega_0,L}({\bf x},{\bf x'};\beta)$ has the following properties:

{\rm i}. $(0,\infty)\ni \beta \longmapsto S_{\omega}(\beta)$ is strongly 
differentiable and $s-\lim_{\beta\searrow 0}S_{\omega}(\beta)={\bf 1}$;

{\rm ii}. ${\rm Ran}S_{\omega}(\beta)\in {\rm Dom}(H_{1,L}(\omega))$ and 
$\frac{\partial}{\partial\beta}S_{\omega}(\beta)f+
H_{1,L}(\omega)S_{\omega}(\beta)f=R_{\omega}(\beta)f$, where 
$R_{\omega}(\beta)$ is given by:
\begin{eqnarray}\label{cincii'''}
 &{}& R_{\omega}({\bf x},{\bf x'};\beta)= 
e^{\imath(\delta\omega)\varphi({\bf x},{\bf x'})}
 \left [ (\delta\omega)^2{\bf a}^2({\bf x}-{\bf x'})
G_{\omega_0,L}({\bf x},{\bf x'};\beta) \right . +\nonumber \\ 
&+& \left . 2 (\delta\omega) {\bf a}({\bf x}-{\bf x'})\cdot (\imath\nabla_{\bf x}
+\omega_0 {\bf a}({\bf x})) 
G_{\omega_0,L}({\bf x},{\bf x'};\beta)\right ].
\end{eqnarray}      
\end{proposition}
{\it Proof.} i. Rewriting the integral kernel of $S_{\omega}(\beta)$ as:
\begin{equation}\label{cinco}
S_{\omega}({\bf x},{\bf x'};\beta)=G_{\omega,L}({\bf x},{\bf x'};\beta)+
\left [ e^{\imath(\delta\omega)\varphi({\bf x},{\bf x'})}-1\right ]
G_{\omega,L}({\bf x},{\bf x'};\beta),
\end{equation}
and using the diamagnetic inequality, one can reproduce the argument from 
(\ref{strlyy}) in order to prove that the second term converges in norm to 
zero. Clearly, the first one converges strongly to one. 

If $\{\psi_j\}$ and $\{E_j\}$ denote the sets of eigenvectors and eigenvalues 
of $H_{1,L}(\omega)$, then:
\begin{equation}\label{cinco'}
G_{\omega,L}({\bf x},{\bf x'};\beta)=\sum_je^{-\beta E_j}\psi_j({\bf x})
\overline{\psi_j}({\bf x'}),
\end{equation}
where the series is absolutely and uniformly convergent on 
$\overline{\Lambda}\times\overline{\Lambda}$. This can be seen from the fact 
that the semigroup is trace class and that the eigenfunctions belong to 
$D_0$ and admit the estimate:
\begin{equation}\label{cinco''}
|\psi_j|({\bf x})\leq {\rm const}(L)\;(E_j+1),
\end{equation}
obtained from the fact that the resolvent $[H_{1,L}(\omega)+1]^{-1}$ is 
bounded between ${\bf L^2}(\Lambda)$ and ${\bf L^\infty}(\Lambda)$. It follows 
that uniformly in $\Lambda$:
\begin{equation}\label{cincoo}
|G_{\omega,L}(\cdot,\cdot;\beta+\delta\beta)-
G_{\omega,L}(\cdot,\cdot;\beta)-\frac{\partial G_{\omega,L}}{\partial\beta}
(\cdot,\cdot;\beta)|\leq {\rm const}(L)\;(\delta\beta)^2,
\end{equation}
which is sufficient for the strong differentiability (see also (\ref{stryu})). 

ii. One has to make the same steps as in the proof of the third point of 
Proposition \ref{pert}. As for the norm of $R(\beta)$, let us see that is 
independent of $L$ and is integrable in zero: 
from (\ref{cincii'''}), (\ref{sase}), (\ref{ineh}) and (\ref{1.11}), one can 
obtain an estimate on the kernel of $R_\omega(\beta)$ of the following form:
\begin{equation}\label{Hids}
|R_\omega({\bf x},{\bf x'};\beta)|\leq c_9\delta\omega
(1+\beta)^s G_{0,\infty}({\bf x},{\bf x'};c_{10}\beta),
\end{equation}
which implies that its $B({\bf L}^2(\Lambda))$ norm is bounded by a constant 
multiplied with $\delta\omega$ (see (\ref{9999})).

\vspace{0.5cm}

We shall give now without proof a result which gives sufficient conditions for 
an operator defined in $B({\bf L}^2(\Lambda))$ to be trace class:
\begin{proposition}\label{lema22}
Let $\{T_n\}$ a sequence of trace class operators, converging to $T$ in 
$B({\bf L}^2(\Lambda))$. If $\sup_n||T_n||_{{\cal I}_1}\leq c<\infty$, then 
$T\in {\cal I}_1$ and $||T||_{{\cal I}_1}\leq c$.
\end{proposition}
{\it Remark.} Assume that 
an operator $T$ is defined by a $B({\bf L}^2(\Lambda))$-norm 
Riemann integral on the interval 
$[a,b]$, with a continuous trace class integrand $S(t)$. If 
$\int_a^bdt\;||S(t)||_{{\cal I}_1}\leq c<\infty$, then $T$ is trace class and
$$||T||_{{\cal I}_1}\leq c,\quad {\rm tr}\;T=\int_a^bdt\;{\rm tr}\;S(t).$$

Denote with $\tilde{R}(\omega,\beta)$ the bounded operator given by the kernel:
\begin{equation}\label{restul}
\tilde{R}({\bf x},{\bf x'};\beta)=2\:{\bf a}({\bf x}-{\bf x'})
\cdot (-\imath\nabla -\omega_0 {\bf a}({\bf x}))
G_{\omega_0,L}({\bf x},{\bf x'};\beta).
\end{equation}
Among other things, the next proposition proves (\ref{heurtu'}):
\begin{proposition}\label{lema} Take $\beta >0$ and 
$\omega=\omega_0+\delta\omega\in\Omega$. 

{\rm i}. The operator $S_\omega(\beta)$ is trace class and moreover, there 
exists a positive numerical constant $c$ such that:
\begin{equation}\label{central1}
||W_L(\beta,\omega)
-S_\omega(\beta)||_{B({\bf L}^2(\Lambda))}\leq 
c\;\delta\omega\;\quad and
\end{equation}
\begin{equation}\label{central2}
||W_L(\beta,\omega)
-S_\omega(\beta)||_{{\cal I}_1}\leq 
c\;\delta\omega\;L^3.
\end{equation}

ii. For all ${\bf x},{\bf x'}\in \Lambda$ and uniformly in $L$, one has:
\begin{eqnarray}\label{central}
&{}&  G_{\omega,L}({\bf x},{\bf x'};\beta) = 
e^{\imath\delta\omega\varphi({\bf x},{\bf x'})}
G_{\omega_0,L}({\bf x},{\bf x'};\beta)+ \\
&+& \delta\omega\int_0^\beta d\tau\int_{\Lambda}d{\bf y}\;
e^{\imath\delta\omega\varphi({\bf x},{\bf y})}
G_{\omega_0,L}({\bf x},{\bf y};\beta-\tau)
e^{\imath\delta\omega\varphi({\bf y},{\bf x'})}
\tilde{R}({\bf y},{\bf x'};\tau)
+ {\cal O}((\delta\omega)^2).\nonumber
\end{eqnarray}
\end{proposition}
{\it Proof.} i. We know that as bounded operators:
\begin{equation}\label{HillepR}
W_L(\beta,\omega)=
S_\omega (\beta)-\int_0^\beta d\tau\;
W_L(\beta-\tau,\omega)R_\omega(\tau).
\end{equation}

We have already seen that the $B({\bf L}^2(\Lambda))$ norm of $R_\omega(\tau)$ 
is bounded by a constant 
multiplied with $\delta\omega$ (see (\ref{Hids})). Its Hilbert-Schmidt 
norm is bounded by (see (\ref{9999'''}):
\begin{equation}\label{Hilbsc}
||R_\omega(\tau)||_{{\cal I}_2}\leq {\rm const}\;\delta\omega\;
\frac{(1+\tau)^s}{\tau^{3/4}} L^{3/2}.
\end{equation}
For the semigroup, we know that $||W_L||_{B({\bf L}^2(\Lambda))}\leq 1$ and 
from (\ref{1.11}) and (\ref{9999'''}) it follows that:
\begin{equation}\label{Hilbsc'}
||W_L(\beta-\tau,\omega) ||_{{\cal I}_2}\leq 
{\rm const} \frac{1}{(\beta -\tau)^{3/4}} L^{3/2}.
\end{equation}
With the help of the well known inequality 
$||A\;B||_{{\cal I}_1}\leq ||A||_{{\cal I}_2}
||B||_{{\cal I}_2}$, it follows that in both situations ($B({\bf L}^2)$ and 
${\cal I}_1({\bf L}^2)$) the singularities in 
$\tau$ are integrable (see the previous remark), and the desired bounds follow 
easily.

ii. The formula (\ref{central}) is obtained from (\ref{HillepR}) by isolating 
the term which contains $\delta\omega$.

\vspace{0.5cm}

The next proposition imposes sufficient conditions on a trace class 
integral operator such that its trace to be equal to the integral of the 
kernel's diagonal (see [R-S 1]):
\begin{proposition}\label{trasda}
Let $T\in{\cal I}_1({\bf L}^2(\Lambda))$, given by the integral kernel 
$T({\bf x},{\bf x'})\in C(\overline{\Lambda}\times\overline{\Lambda})$. Then:
\begin{equation}\label{greca}
{\rm tr}\;T=\int_\Lambda d{\bf x}\;T({\bf x},{\bf x}).
\end{equation}
\end{proposition}

The rest of this subsection is dedicated to the proof of (\ref{heurtu}). 
Fix $\beta,\tau >0.$ 
Let $0<\omega_0 <\omega_1$ and let $\Omega =[\omega_0,\omega_1]$. Clearly, 
$g_\omega(\xi,z;\beta,\tau)$ is trace class and admits a continuous integral 
kernel given by the following series, which is 
absolutely and uniformly convergent on $\Lambda\times\Lambda$ (see also 
(\ref{cinco'})):
\begin{equation}\label{gbetatau'}
T_\omega({\bf x},{\bf x'})=\sum_{j}
[\xi -z\exp{(-\beta E_j)}]^{-1}z\exp{(-\tau E_j)}\psi_j({\bf x})
\overline{\psi_j}({\bf x'}),
\end{equation}
Notice that in order to simplify the notations, we did not specify 
the dependence on $\xi$, $z$, $\beta$ and $\tau$.

Let us start with the equation satisfied by $T_\omega({\bf x},{\bf x'})$:
\begin{proposition}\label{xxxc}
As continuous functions:
\begin{eqnarray}\label{ecintg}
\xi T_\omega({\bf x},{\bf x'})-z\int_\Lambda d{\bf y}\;
G_{\omega,L}({\bf x},{\bf y};\beta) T_\omega({\bf y},{\bf x'})=
zG_{\omega,L}({\bf x},{\bf x'};\tau).
\end{eqnarray}
\end{proposition}
{\it Proof.} The above equality 
is nothing but the rewriting in terms of integral 
kernels of an identity between bounded operators:
\begin{equation}\label{gbetatau''}
[\xi -zW_L(\beta,\omega)]g_\omega(\xi,z;\beta,\tau)=
zW_L(\tau,\omega).
\end{equation}

\vspace{0.5cm}

For further purposes, we shall prove that for all $L\geq 1$, 
$|T_\omega({\bf x},{\bf x'})|\sim e^{-\alpha
|{\bf x}-{\bf x'}|}$ for some positive $\alpha$. We need first 
a few definitions: let $\rho ({\bf x}):=(1+{\bf x}^2)^{1/2}$ and 
$\alpha\geq 0$. It is known that the partial 
derivatives up to the second order 
of $\rho$ are bounded by a numerical constant and moreover:
\begin{equation}\label{rho}
\forall {\bf x}\in {\RM}^3,\: e^{\pm \alpha\rho ({\bf x})}
 e^{\mp\alpha |{\bf x}|}\leq {\rm const}(\alpha).
\end{equation}
Fix ${\bf x}_0\in\Lambda$. Denote with $A(\alpha)$ the multiplication operator 
with $e^{-\alpha\rho (\cdot -{\bf x}_0)}$. Then $A(\alpha)$ and 
$A^{-1}(\alpha)=A(-\alpha)$ are bounded operators and invariate $D_0$ (see 
\ref{oreir})). An useful result is contained in the next proposition:
\begin{proposition}\label{Ae1}
 
{\rm i.} The operator $A(\alpha)W_L(\tau,\omega)A(-\alpha)$ belongs to 
$B({\bf L}^2)$, and has a norm which is uniformly bounded in $L$, 
 ${\bf x}_0\in \Lambda$, $\omega\in\Omega$ and $0<\tau \leq\beta$;

{\rm ii.} The operator  $A(\alpha)W_L(\beta,\omega)A(-\alpha)$ belongs to 
$B({\bf L}^2,{\bf L}^\infty)$, having a norm which is uniformly bounded in 
$L$, $\omega$ and ${\bf x}_0$.
\end{proposition}
{\it Proof.} i. The inequality (\ref{rho}) allows us to replace $A$ with the 
multiplication operator given by $ e^{\alpha |{\bf x}-{\bf x}_0|}$. Let 
$\psi\in {\bf L}^2(\Lambda)$ and define: 
\begin{equation}\label{linfty}
\phi :=A(\alpha)W_L(\tau,\omega)A(-\alpha)=\int_\Lambda d{\bf y}\;
e^{-\alpha\rho ({\cdot} -{\bf x}_0)}G_{L,\omega}({\cdot},{\bf y};\tau)
e^{\alpha\rho ({\bf y} -{\bf x}_0)}\psi({\bf y}).
\end{equation}
Let us remark an elementary estimate, which is true for all 
$0<\tau \leq\beta $: 
\begin{equation}\label{linfty'}
e^{\alpha |{\bf x}-{\bf x'}|}\exp{\left [-\frac{|{\bf x}-{\bf x'}|^2}{4\tau}
\right ]}\leq {\rm const}(\alpha,\beta).
\end{equation}
Applying (\ref{rho}), the triangle inequality, (\ref{1.11}) and 
(\ref{linfty'}) one obtains:
\begin{eqnarray}\label{linfty1}
|\phi |({\bf x}) &\leq & {\rm const}(\alpha)\int_{\Lambda}d{\bf x'}\;
e^{\alpha |{\bf x}-{\bf x'}|}
|G_{\omega,L}({\bf x},{\bf x'};\tau)|\;|\psi|({\bf x'})\leq \nonumber \\
&\leq& {\rm const}(\alpha ,\beta)\int_{\Lambda}d{\bf x'}\;
G_{0,\infty}({\bf x},{\bf x'};2\tau)|\psi|({\bf x'}).
\end{eqnarray}
The result follows from (\ref{9999''}) and (\ref{9999}).

ii. We apply the Schwartz inequality in (\ref{linfty1}) with $\tau =\beta$ and 
then use (\ref{9999'''}).

\vspace{0.5cm}

\begin{proposition}\label{Ae2}
Under the same conditions as above, there exists a sufficiently small 
$0<\alpha <1$ such that the following inequalities are true in 
$B({\bf L}^2(\Lambda))$, uniformly in $L$, $\omega$ and ${\bf x}_0$:

{\rm i.} $||W_L(\beta,\omega)-
A(\alpha)W_L(\beta,\omega)A(-\alpha)||\leq \alpha\;
{\rm const}(\beta)$;

{\rm ii.} Uniformly in ${\xi\in {\cal C}}$ and ${z\in K}$ one has:
$$||A(\alpha)
[\xi -zW_L(\beta,\omega)]^{-1}A(-\alpha)
||_{B({\bf L}^2(\Lambda))}
\leq {\rm const}(\beta).
$$
\end{proposition}
{\it Proof.} i. Let $S(\beta)=A(\alpha)W_L(\beta,\omega)A(-\alpha)$. We will 
see that $S(\beta)$ obeys the conditions of Proposition \ref{general1}. 
From Proposition \ref{Ae1} follows condition A. Then $S(\beta)$ is 
strongly differentiable, has its range included in the domain of 
$ H_{1,L}(\omega)$ and converges strongly to one. Define 
$B:=H_{1,L}(\omega)-A(\alpha)H_{1,L}(\omega)A(-\alpha)$, or in other form:
\begin{equation}
B=2\imath\alpha ({\bf p}-\omega {\bf a})\cdot \nabla\rho(\cdot-{\bf x}_0) 
+\alpha (\Delta\rho)(\cdot-{\bf x}_0)-\alpha^2 |\nabla\rho(\cdot-{\bf x}_0)|^2.
\end{equation}
A well known result says (see [S]) that 
$\left\Arrowvert B[H_{1,L}(\omega)+1]^{-1/2}\right\Arrowvert \leq {\rm const}$.

By direct computation, 
$R(\tau)=B\;A(\alpha)W_L(\tau,\omega)A(-\alpha)$; a rough estimate gives 
$||R(\tau)||\leq {\rm const}(L)/\sqrt{\tau}$ and even if the constant 
behaves badly with $L$, the norm is integrable in zero with respect to $\tau$. 

In conclusion:
\begin{eqnarray}\label{crab1}
W_L(\beta,\omega) -S(\beta) &=& -\int_0^\beta d\tau\;
W_L(\beta-\tau,\omega)R(\tau)\nonumber \\
&=& -\int_0^\beta d\tau\;W_L(\beta-\tau,\omega)
BS(\tau),
\end{eqnarray}
where the integral converges in norm. But uniformly in $L$ and ${\bf x}_0$ 
there exists a numerical constant such that:
\begin{equation}\label{crab}
||W_L(\beta-\tau,\omega)B||\leq \alpha \frac{{\rm const}}{
\sqrt{\beta -\tau}}.
\end{equation}
From (\ref{crab}), (\ref{crab1}) and point i of Proposition \ref{Ae1}, the 
needed estimate follows. 

ii. Using point i, the estimate (\ref{K1}) and the identity ($\alpha$ small 
enough):
\begin{eqnarray}
&{}&  A(\alpha)
[\xi -zW_L(\beta,\omega)]^{-1}A(-\alpha) =
[\xi -zA(\alpha)W_L(\beta,\omega)A(-\alpha)]^{-1}=\nonumber \\
&=& \sum_{j\geq 0}[\xi -zW_L(\beta,\omega)]^{-1}z^j\cdot \nonumber \\
&\cdot& \left \{[A(\alpha)W_L(\beta,\omega)A(-\alpha)-W_L(\beta,\omega)]
[\xi -zW_L(\beta,\omega)]^{-1}\right \}^j,\nonumber
\end{eqnarray}
the result follows.

\vspace{0.3cm}

\begin{corollary}\label{asd}
The operator $ A(\alpha)g_\omega(\xi,z;\beta,\tau) A(-\alpha)$ belongs to 
$B({\bf L}^2,{\bf L}^\infty)$ if $\alpha$ is small enough, and uniformly in 
$\xi$, $z$, ${\bf x}_0$, $\omega$ and $L$ one has:

{\rm i.} $|| A(\alpha)g_\omega
(\xi,z;\beta,\tau) A(-\alpha)||\leq {\rm const}(\beta,\tau);$

{\rm ii.}  $\int_\Lambda d{\bf y}\;e^{2\alpha |{\bf y}-{\bf x}_0|}
|T_\omega({\bf x}_0,{\bf y})|^2\leq {\rm const}(\beta,\tau);$

{\rm iii.} $ e^{\alpha |{\bf x}-{\bf y}|}
|T_\omega({\bf x},{\bf y})|\leq {\rm const}(\beta,\tau).$
\end{corollary}
{\it Proof.} i. It is an immediate consequence of Propositions 
\ref{Ae1} and \ref{Ae2}.

ii. Let  $\phi = A(\alpha)g_\omega
(\xi,z;\beta,\tau) A(-\alpha)\psi$, where $\phi$ is bounded and continuous. 
From i it follows:
\begin{equation}
|\phi ({\bf x}_0)|=|\int_\Lambda d{\bf y}\;e^{\alpha \rho({\bf y}-{\bf x}_0)}
T_\omega
({\bf x}_0,{\bf y})\psi({\bf y})|\leq {\rm const}(\beta,\tau)
||\psi||_{{\bf L}^2},
\end{equation}
and the result follows 
from the representation theorem of linear and continuous functionals on 
${\bf L}^2$.

iii. Rewrite the identity $g_\omega(\xi,z;\beta,\tau)=g_\omega
(\xi,z;\beta,\tau/2)W_L(\tau/2,\omega)$ in terms of integral kernels, use 
(\ref{rho}), (\ref{1.11}), ii, the triangle and Schwartz inequalities, and 
the proof is completed.

\vspace{0.3cm}

Let $\delta\omega >0$ be such that $\omega=\omega_0 +
\delta\omega\in\Omega$. Define the bounded operator 
$\tilde{g}_{\omega} (\xi,z;\beta,\tau)$ given by the following integral kernel:
\begin{equation}\label{314159}
\tilde{T}_\omega 
({\bf x},{\bf x}'):=e^{\imath\delta\omega\varphi 
({\bf x},{\bf x}')}T_{\omega_0}({\bf x},{\bf x}').
\end{equation}
Equations (\ref{K1}) and (\ref{central1}) imply that if $\delta\omega$ is 
sufficiently small then there exists a numerical constant such that 
uniformly in $L$:
\begin{equation}\label{newman}
\sup_{\xi}\sup_{z}||[\xi -zS_{\omega}(\beta)]^{-1}||\leq {\rm const} 
\end{equation}
and:
\begin{equation}\label{yur}
\sup_{\xi}\sup_{z}||[\xi -zS_{\omega}(\beta)]^{-1}
-[\xi -zW_L(\beta,\omega)]^{-1}
||\leq {\rm const}\;\delta\omega.
\end{equation}
We state now an important property of $\tilde{g}_{\omega}(\xi,z;\beta,\tau)$:
\begin{proposition}\label{qwe}
Under the above conditions, there exists a numerical constant such that if 
$\delta\omega$ is small enough, then uniformly in $\xi$, $z$ and $L$, 
the following $B({\bf L}^2)$ estimate takes place:
\begin{equation}\label{newman'}
||[\xi -zS_{\omega}(\beta)]^{-1}zS_{\omega}(\tau)-
\tilde{g}_{\omega} (\xi,z;\beta,\tau)||\leq {\rm const} \;\delta\omega.
\end{equation}
\end{proposition}
{\it Proof.} The integral kernel of the operator 
$[\xi -zS_{\omega}(\beta)]\tilde{g}_{\omega} (\xi,z;\beta,\tau)$ is given by:
\begin{equation}\label{newman''}
\xi \tilde{T}_{\omega}({\bf x},{\bf x'})-z\int_\Lambda d{\bf y}\;
S_{\omega}({\bf x},{\bf y};\beta) \tilde{T}_{\omega}
({\bf y},{\bf x'}).
\end{equation}
Let us notice a crucial property of the magnetic phase:
\begin{equation}\label{newman'''}
\varphi({\bf x},{\bf y})+\varphi({\bf y},{\bf x}')=
\varphi({\bf x},{\bf x}')+{\rm fl}({\bf x},{\bf y},{\bf x}'),
\end{equation}
where ${\rm fl}({\bf x},{\bf y},{\bf x}')=1/2\;{\bf B}\cdot
[({\bf y}-{\bf x}')\wedge ({\bf x}-{\bf y})]$. Using (\ref{newman'''}) and 
(\ref{ecintg}) in (\ref{newman''}) we obtain:
\begin{equation}\label{redf}
[\xi -zS_{\omega}(\beta)]\tilde{g}_{\omega} (\xi,z;\beta,\tau)=
zS_{\omega}(\tau)+R,
\end{equation}
where $R$ is an integral operator given by: 
\begin{equation}
-z\;e^{\imath\delta\omega\varphi 
({\bf x},{\bf x}')}\int_\Lambda d{\bf y}\;(e^{\imath\delta\omega\;
{\rm fl}({\bf x},{\bf y},{\bf x}')}-1)\;
G_{\omega_0,L}({\bf x},{\bf y};\beta) T_{\omega_0}
({\bf y},{\bf x'}).
\end{equation}
Because $|e^{\imath\delta\omega\;
{\rm fl}({\bf x},{\bf y},{\bf x}')}-1|\leq\delta\omega |{\bf x}-{\bf y}|\;
|{\bf y}-{\bf x}'|$, denoting with $P$ the operator given by 
$|{\bf x}-{\bf y}|\;|zG_{\omega_0,L}({\bf x},{\bf y};\beta)|$ and with 
$Q$ the operator corresponding to 
$|{\bf y}-{\bf x}'|\;| T_{\omega_0}
({\bf y},{\bf x'})|$, it follows that $||R||\leq\delta\omega||P||\;||Q||$. 
Using (\ref{1.11}), Corollary \ref{asd} iii., (\ref{9999}) and 
(\ref{newman}), the proof is completed.

\vspace{0.5cm}

Employing 
(\ref{central1}), (\ref{newman'}), (\ref{yur}) and (\ref{newman}) in 
the next equality:
\begin{eqnarray}
&{}& g_\omega(\xi,z;\beta,\beta/2)-\tilde{g}_\omega(\xi,z;\beta,\beta/2)=\nonumber \\ 
&=& [(\xi-zW_L(\beta,\omega))^{-1}-(\xi-zS_\omega(\beta))^{-1}]zW_L(\beta/2,\omega)
+\nonumber \\
&+& (\xi-zS_\omega(\beta))^{-1}z[W_L(\beta/2,\omega)-S_\omega(\beta/2)]+\nonumber \\
&+& (\xi-zS_\omega(\beta))^{-1}zS_\omega(\beta/2)
 -\tilde{g}_\omega(\xi,z;\beta,\beta/2),
\end{eqnarray}
(\ref{heurtu}) is straightforward. 

Let us end this subsection by proving that the operator 
$$A_\omega(\xi,z;\beta)=\tilde{g}_\omega(\xi,z;\beta,\beta/2)
S_\omega(\beta/2)$$ has the same trace as 
$g_{\omega_0}(\xi,z;\beta,\beta)$. Indeed, because $A_\omega$ fulfills the 
conditions of Proposition \ref{trasda} and noticing that $\varphi({\bf x},
{\bf x'})=-\varphi({\bf x'},{\bf x})$, one can write:
\begin{eqnarray}
{\rm tr}A_\omega(\xi,z;\beta) &=& \int_{\Lambda^2}d{\bf x}\;d{\bf x'}\;
T_{\omega_0}(\xi,z;\beta,\beta/2;
{\bf x},{\bf x'})G_{\omega_0,L}({\bf x'},{\bf x};\beta/2)\nonumber \\
&=& \int_{\Lambda}d{\bf x}\;T_{\omega_0}(\xi,z;\beta,\beta;
{\bf x},{\bf x})={\rm tr}g_{\omega_0}(\xi,z;\beta,\beta).
\end{eqnarray}

\subsection{The proof of II and III}

The analyticity of $\Gamma_\infty(\beta,z,\omega_0)$ in ${\cal D}$ follows 
from the bound (see (\ref{gbosonic})) $|g_\sigma(\zeta)|\leq {\rm const}(
\sigma,K)|\zeta|$ where $K$ is some compact in 
${\CM}\setminus [1,\infty)$. 

In what follows, we will prove that if $z\in {\cal D}_0:=\{|z|<1\}$, then:
\begin{equation}\label{poiuytrew}
\lim_{L\rightarrow\infty}\Gamma_L(\beta,z,\omega_0)=
\Gamma_\infty(\beta,z,\omega_0).
\end{equation}

Because $|z|<1$, the grand canonical pressure will be (see (\ref{1.166})):
\begin{equation}\label{vespuci2}
P_L(\beta,z,\omega)=\sum_{n=1}^{\infty}
\frac{z^n}{n}\;\frac{1}{\beta L^3}\;{\rm tr}W_L(n\beta,\omega)=
\sum_{n=1}^{\infty}
\frac{z^n}{n}\left ( \frac{1}{\beta L^3}\int_{\Lambda}d{\bf x}\:
G_{\omega,L}({\bf x},{\bf x};n\beta)\right ).
\end{equation}
Under the same conditions, the magnetization reads as:
\begin{equation}\label{gussi}
\Gamma_L(\beta,z,\omega_0)=-\frac{e}{c}\sum_{n=1}^{\infty}\frac{z^n}{n}\;
\frac{1}{\beta L^3}\;{\rm tr}\left [\frac{\partial W_L}{\partial\omega}\right ]
(n\beta,\omega_0).
\end{equation}
An important quantity is the integral kernel of the semigroup defined on the 
whole space:
\begin{eqnarray}\label{sapte}
&{}& G_{\omega,\infty}({\bf x},{\bf x'};\beta)= 
\frac{e^{\imath\omega\varphi
({\bf x},{\bf x'})}}
{(2\pi\beta)^{3/2}}\;
\frac{\omega\beta /2}{\sinh {(\omega\beta /2)}}\cdot \\ 
&\cdot & \exp{\left \{
-\frac{1}{2\beta}\left [\frac{\omega\beta /2}{\tanh {(\omega\beta /2)}}\;
({\bf e_3}\wedge ({\bf x}- {\bf x'}))^2 +
({\bf e_3}\cdot ({\bf x}- {\bf x'}))^2
\right ]
\right \}}.
\end{eqnarray} 
Denote with:
$$g(\beta,\omega)=G_{\omega,\infty}({\bf x},{\bf x};\beta)=\frac{1}
{(2\pi\beta)^{3/2}}\;\frac{\omega\beta /2}{\sinh {(\omega\beta /2)}}.$$
Then it is easy to see that if $|z|<1$: 
$$P_\infty(\beta,z,\omega_0)=
\sum_{n=1}^{\infty}
\frac{z^n}{n}\frac{g(n\beta,\omega_0)}{\beta},\quad 
\Gamma_\infty(\beta,z,\omega_0)=-\frac{e}{\beta c}\sum_{n=1}^{\infty}
\frac{z^n}{n}\frac{\partial g}{\partial\omega}(n\beta,\omega_0).$$
One of the results in [M-M-P 1] can be adapted to our problem and gives:
\begin{equation}\label{gussi1}
\lim_{L\rightarrow\infty}
\frac{1}{L^3}\;{\rm tr}\left [\frac{\partial W_L}{\partial\omega}\right ]
(\beta,\omega_0)=\frac{\partial g}{\partial\omega}(\beta,\omega_0).
\end{equation}
If we prove the existence of a positive 
function $f$ with at most polynomial growth such that:
\begin{equation}\label{vespuci3}
\left \arrowvert\frac{1}{ L^3}\;
{\rm tr}\left [\frac{\partial W_L}{\partial\omega}\right ]\right\arrowvert
(n\beta,\omega_0)\leq f(n\beta),
\end{equation} 
then (\ref{poiuytrew}) would be true.

The next corollary is a direct consequence of Proposition \ref{lema} ii.:

\begin{corollary}\label{corolar}
Under the conditions of Proposition \ref{lema} one has:

\begin{eqnarray}\label{derivata} 
&{}&\lim_{\delta\omega\searrow 0}\frac{1}{\delta\omega}
\int_\Lambda d{\bf x}\;
[G_{\omega_0+\delta\omega,L}({\bf x},{\bf x};\beta)-
G_{\omega_0,L}({\bf x},{\bf x};\beta)]=  \\
&=& {\rm tr}\left [\frac{\partial W_L}{\partial\omega}\right ](\beta,\omega_0)=
2\int_0^\beta d\tau \int_{\Lambda^2}d{\bf x}\;d{\bf y}\cdot \nonumber \\
&\cdot & G_{\omega_0,L}({\bf x},{\bf y};\beta-\tau)
{\bf a}({\bf y}-{\bf x})
\cdot (-\imath\nabla_{\bf y} -\omega_0 {\bf a}({\bf y}))
G_{\omega_0,L}({\bf y},{\bf x};\tau).\nonumber
\end{eqnarray}
\end{corollary}
\vspace{0.5cm}

Use (\ref{1.11}), (\ref{sase}), (\ref{ineh}) and (\ref{9999'}) in 
(\ref{derivata}) and (\ref{vespuci3}) follows. The proof of (\ref{poiuytrew}) 
is now completed. 
\vspace{0.5cm} 

\noindent 
{\it Remarks.} 1. What can we say about the same problem for Fermi particles 
(say electrons, where $H_{1,L}(\omega)$ should be replaced with 
the Pauli operator)? Knowing 
that in this case $\inf\sigma(H_{1,\infty}(\omega))=0$ 
(i.e. is independent of 
$\omega$), 
the grand canonical result (Lemma \ref{limgama}) can be easily restated 
in terms of Fermi statistics: the only thing that changes is the domain on 
which the limit takes place i.e. 
${\CM}\setminus (-\infty, -1]$. As for Theorem 
\ref{magcano}, its proof was based on the fact that there exists a 
compact $K\subset {\CM}\setminus [e^{\beta\omega/2}, \infty)$ which contains 
the circle centered in 
the origin with radius equal to $x_L(\beta,\rho,\omega)$, for all $L\geq L_0$ 
and for all strictly positive $\beta$, $\omega$ and $\rho$. For Fermi 
particles, if one fixes $\rho$ and $\omega$ but makes $\beta$ very large 
(lowers the temperature), then  $x_L(\beta,\rho,\omega)$ 
would be a very large positive quantity, therefore the above circle could 
intersect the negative cut. Of course, if the gas is diluted (say $\rho$ and 
$\omega$ fixed and $\beta$ small), then it could happen that 
$x_\infty(\beta,\rho,\omega)<1$ which means that the circle never intersects 
the cut when $L\geq L_0$, therefore a similar proof can be provided. 
Our conclusion is that the extension of Theorem \ref{magcano} to a Fermi gas 
at low temperature is not trivial and remains an interesting problem. 

2. What about the higher derivatives with respect to $\omega$ (the 
susceptibility for example) at $\omega_0\neq 0$? 
This also remains an open problem, even for the grand 
canonical ensemble. Nevertheless, we think that our approach (the modified 
perturbation theory for Gibbs semigroups) could provide an answer to it.

\vspace{0.5cm}
 
\begin{acknowledgement}
Part of this work was done during a visit to 
Centre de Physique Th\'eorique in Marseille, at the invitation of Professor 
P. Duclos. The financial support of CNCSU grant 13(C) is hereby gratefully 
acknowledged. 
Finally, the author wishes to thank Professors N. Angelescu, M. Bundaru and 
G. Nenciu for their encouragement and fruitful discussions.
\end{acknowledgement}


\begin{thebibliography}{9999999}
\bibitem[A]{d} Angelescu, N.: Ph.D thesis, I.F.A., Bucharest, 1976
\bibitem[A-B-N 1]{abn} Angelescu, N., Bundaru, M., Nenciu, G.: 
On the perturbation of Gibbs semigroups. 
 Commun. Math. Phys. {\bf 42}, 29-30 (1975)
\bibitem[A-B-N 2]{abn'} Angelescu, N., Bundaru, M., Nenciu, G.: 
On the Landau diamagnetism. 
 Commun. Math. Phys. {\bf 42}, 9-28 (1975)
\bibitem[A-C]{ac} Angelescu, N., Corciovei, A.: On free quantum gases in a 
homogeneous magnetic field. Rev. Roum. Phys. {\bf 20}, 661-671 (1975)
\bibitem[B-H-L]{bhl} Broderix, K., Hundertmark, D., Leschke, H.:
Continuity properties of Schrodinger semigroups with magnetic fields. 
Mathematical-Physics preprint archive of University of Texas at Austin.
\bibitem[C-N]{f} Cornean, H.D., Nenciu, G.: On eigenfunction decay for two 
dimensional 
magnetic Schr\"{o}dinger operators. 
Commun. Math. Phys. {\bf 192}, 671-685 (1998)
\bibitem[H]{huang} Huang, K.: Statistical mechanics. New York-London: 
John Wiley \& Sons, Inc., 1963 
\bibitem[H-P]{hp} Hille, E., Phillips, R.S.: Functional integral and 
semigroups. Providence; RI: Am. Math. Soc., 1957
\bibitem[K-U-Z]{ka} Kac, M.,  Uhlenbeck, G.E., Ziff, R.M.: 
The ideal Bose-Einstein gas, 
revisited. Phys. Rep. {\bf 32C}, $n^o$ 4, 169-248 (1977)
\bibitem[K]{ku} Kunz, H.: Surface orbital magnetism. 
J. Stat. Phys. {\bf 76}, 183-207 (1994)
\bibitem[M-M-P 1]{mmpa} Macris, N., Martin, Ph.A., Pul\'e, J.V.: 
Diamagnetic currents. 
Commun. Math. Phys. {\bf 117}, 215-241 (1988)
\bibitem[M-M-P 2]{mmp} Macris, N., Martin, Ph.A., Pul\'e, J.V.: 
Large volume asymptotics of Brownian integrals and orbital magnetism. 
Ann. I.H.P. Phys. Theor. {\bf 66}, 147-183 (1997)
\bibitem[R-S 1,4]{z} 
Reed, M., Simon, B.: Methods of modern mathematical physics 
{\bf I, IV}. New York: Academic Press, 1975
\bibitem[S]{fgs} Simon, B.: Schr\"odinger semigroups. Bull. Am. Math. Soc. 
(N. S.) {\bf 7}, 447-510 (1982)
\end{thebibliography}
\end{document}